\documentclass[twocolumn]{aastex631}

\usepackage{apjfonts}
\usepackage{rotating}
\usepackage[dvipsnames]{xcolor}
\usepackage{amsmath,color,graphicx,tablefootnote,enumitem}
\usepackage{graphics,subfigure,hyperref,float,multirow}
\usepackage[rightcaption]{sidecap}

\newcommand{\CH}[1]{\colhead{#1}}
\newcommand{\D}{$^{\dagger}$}

\newcommand{\W}{$\lambda$}
\usepackage{rotating}

\newcommand{\ebv}{{0.052}}
\newcommand{\ebveu}{{0.110}}
\newcommand{\ebved}{{0.052}}
\newcommand{\FHaCorr}{{$1.985\times10^{-17}$}} 
\newcommand{\FHAHB}{{3.48}}
\newcommand{\FHBHG}{{2.16}}
\newcommand{\SFR}{{3.2}} 
\newcommand{\HAexc}{{0.204}}
\newcommand{\EWHA}{{1457}}
\newcommand{\LHA}{{$1.20\times10^{42}$}}  
\newcommand{\HeIIv}{{530}}

\newcommand{\thigh}{{1.97}}
\newcommand{\thighe}{{0.03}}
\newcommand{\nNIV}{{2.65}} 
\newcommand{\nNIVeu}{{0.39}} 
\newcommand{\nNIVed}{{0.37}} 
\newcommand{\nCIII}{{7.94}} 
\newcommand{\nSiIII}{{4.77}} 
\newcommand{\nSII}{{1.15}} 

\newcommand{\loh}{{7.753}} 
\newcommand{\lohe}{{0.025}} 
\newcommand{\lui}{{-1.24}} 
\newcommand{\luh}{{-0.69}} 

\newcommand{\lnoave}{{-0.391}} 
\newcommand{\lnoavee}{{0.037}}

\shorttitle{GLIMPSE WN}
\shortauthors{Berg et al.}

\begin{document}

\shortauthors{Berg et al.}
\title{A Fleeting GLIMPSE of N/O Enrichment at Cosmic Dawn: 
Evidence for Wolf Rayet N Stars in a $z=6.1$ Galaxy}
\author[0000-0002-4153-053X]{Danielle A. Berg}
\affiliation{Department of Astronomy, The University of Texas at Austin, 2515 Speedway, Stop C1400, Austin, TX 78712, USA}
\affiliation{Cosmic Frontier Center, The University of Texas at Austin, Austin, TX 78712, USA} 

\author[0000-0003-3997-5705]{Rohan P. Naidu}
\affiliation{MIT Kavli Institute for Astrophysics and Space Research, 70 Vassar Street, Cambridge, MA 02139, USA}

 \author[0000-0002-0302-2577]{John Chisholm}
\affiliation{Department of Astronomy, The University of Texas at Austin, 2515 Speedway, Stop C1400, Austin, TX 78712, USA}
\affiliation{Cosmic Frontier Center, The University of Texas at Austin, Austin, TX 78712, USA}

\author[0000-0002-7570-0824]{Hakim Atek}
\affiliation{Institut d'Astrophysique de Paris, CNRS, Sorbonne Universit\'e, 98bis Boulevard Arago, 75014, Paris, France}

\author[0000-0001-7201-5066]{Seiji Fujimoto}
\affiliation{David A. Dunlap Department of Astronomy and Astrophysics, and Department of Physics, 60 St George St, University of Toronto, Toronto, ON M5S 3H8, Canada}

\author[0000-0002-0302-2577]{Vasily Kokorev}
\affiliation{Department of Astronomy, The University of Texas at Austin, 2515 Speedway, Stop C1400, Austin, TX 78712, USA}
\affiliation{Cosmic Frontier Center, The University of Texas at Austin, Austin, TX 78712, USA}

\author[0000-0001-6278-032X]{Lukas J. Furtak}
\affiliation{Department of Astronomy, The University of Texas at Austin, 2515 Speedway, Stop C1400, Austin, TX 78712, USA}
\affiliation{Cosmic Frontier Center, The University of Texas at Austin, Austin, TX 78712, USA}

\author[0000-0002-4343-0487]{Chiaki Kobayashi}
\affiliation{Centre for Astrophysics Research, Department of Physics, Astronomy and Mathematics, University of Hertfordshire, College Lane, Hatfield AL10 9AB, UK}

\author[0000-0001-7144-7182]{Daniel Schaerer}
\affiliation{Observatoire de Genève, Université de Genève, Chemin Pegasi 51, 1290 Versoix, Switzerland}
\affiliation{CNRS, IRAP, 14 Avenue E. Belin, 31400 Toulouse, France}

\author[0000-0002-8192-8091]{Angela Adamo}
\affiliation{Department of Astronomy, Oskar Klein center, Stockholm University, AlbaNova University center, SE-106 91 Stockholm, Sweden}

\author[0000-0001-7232-5355]{Qinyue Fei}
\affiliation{David A. Dunlap Department of Astronomy and Astrophysics, University of Toronto, 50 St. George Street, Toronto, Ontario, M5S 3H4, Canada}

\author[0000-0002-3897-6856]{Damien Korber}
\affiliation{Observatoire de Genève, Université de Genève, Chemin Pegasi 51, 1290 Versoix, Switzerland}

\author[0000-0003-2871-127X]{Jorryt Matthee}  \affiliation{Institute of Science and Technology Austria (ISTA), Am Campus 1, 3400 Klosterneuburg, Austria}

\author[0000-0001-8442-1846]{Rui Marques-Chaves} 
\affiliation{Observatoire de Genève, Université de Genève, Chemin Pegasi 51, 1290 Versoix, Switzerland}

\author[0009-0000-2997-7630]{Zorayda Martinez}
\affiliation{Department of Astronomy, The University of Texas at Austin, 2515 Speedway, Stop C1400, Austin, TX 78712, USA}
\affiliation{Cosmic Frontier Center, The University of Texas at Austin, Austin, TX 78712, USA} 

\author[0000-0001-5538-2614]{Kristen.~B.~W. McQuinn}
\affiliation{Space Telescope Science Institute, 3700 San Martin Drive, Baltimore, MD, 21218, USA}
\affiliation{Rutgers University, Department of Physics and Astronomy, 136 Frelinghuysen Road, Piscataway, NJ 08854, USA} 

 \author[0000-0002-8984-0465]{Julian B.~Mu\~noz}
\affiliation{Department of Astronomy, The University of Texas at Austin, 2515 Speedway, Stop C1400, Austin, TX 78712, USA}
\affiliation{Cosmic Frontier Center, The University of Texas at Austin, Austin, TX 78712, USA}

\author[0000-0001-5851-6649]{Pascal A. Oesch}
\affiliation{Observatoire de Genève, Université de Genève, Chemin Pegasi 51, 1290 Versoix, Switzerland}
\affiliation{Cosmic DAWN Center, Niels Bohr Institute, University of Copenhagen, Jagtvej 128, K\o benhavn N, DK-2200, Denmark}

\author[0000-0001-8419-3062]{Alberto Saldana-Lopez}
\affiliation{Department of Astronomy, Oskar Klein Centre, Stockholm University, 106 91 Stockholm, Sweden}

\author[0000-0001-6106-5172]{Daniel P. Stark}
\affiliation{Department of Astronomy, University of California, Berkeley, Berkeley, CA 94720, USA}

\author[0000-0003-4717-0376]{Mabel G. Stephenson}
\affiliation{Department of Astronomy, The University of Texas at Austin, 2515 Speedway, Stop C1400, Austin, TX 78712, USA}
\affiliation{Cosmic Frontier Center, The University of Texas at Austin, Austin, TX 78712, USA}

\author[0000-0003-4512-8705]{Tiger Yu-Yang Hsiao} 
\affiliation{Department of Astronomy, The University of Texas at Austin, 2515 Speedway, Stop C1400, Austin, TX 78712, USA}
\affiliation{Cosmic Frontier Center, The University of Texas at Austin, Austin, TX 78712, USA}


\begin{abstract}
We present the discovery of extreme nitrogen enrichment by Wolf Rayet 
nitrogen (WN) stars in the metal-poor ($\sim10\%Z_\odot$), lensed, 
compact ($R_{\rm eff}\sim20$ pc) galaxy RXCJ2248 at $z=6.1$, revealed 
by unprecedentedly deep JWST/NIRSpec medium-resolution spectroscopy 
from the GLIMPSE-D Survey. 
The exquisite signal-to-noise ratio reveals multiple high-ionization nebular lines and 
broad Balmer and [\ion{O}{3}] components (FWHM$\sim700-3000$ km s$^{-1}$). 
We detect broadened \ion{He}{2} \W1640 and \W4687 (FWHM$\sim\HeIIv$ km s$^{-1}$) 
and strong \ion{N}{3} \W4642 emission consistent with a population of 
WN stars, making RXCJ2248 the most distant galaxy with confirmed WR features to date. 
We measure the multiphase nebular density across five ions, the direct-method 
metallicity ($12+\log(\rm O/H)=\loh\pm\lohe$), and a nonuniform elemental enrichment 
pattern of extreme N/O enhancement ($\log(\rm N/O)=\lnoave\pm\lnoavee$ from N$^+$, 
N$^{+2}$, and N$^{+3}$) but suppressed C/O relative to empirical C/N trends. 
We show that this abundance pattern can be explained by enrichment from a dual-burst 
with a low WR carbon/WN ratio, as expected at low metallicities. 
Crucially, these signatures can only arise during a brief, rare evolutionary window 
shortly after a burst ($\sim3-6$ Myr), when WN stars dominate chemical feedback but 
before dilution by later yields (e.g., supernovae). 
The observed frequency of strong N emitters at high$-z$ implies a $\sim50$ Myr burst 
duty cycle, suggesting that N/O outliers may represent a brief but ubiquitous phase 
in the evolution of highly star-forming early galaxies. 
The WN detection in RXCJ2248, therefore, provides the first direct evidence of 
WR-driven nitrogen enrichment in the first billion years of the Universe and 
a novel timing argument for the bursty star formation cycles that shaped 
galaxies at cosmic dawn.
\end{abstract}

\keywords{Chemical abundances (224), H II regions (694), Interstellar medium (847), Interstellar line emission (844), High-redshift galaxies (734), Star-forming galaxies (1560), Galaxy chemical evolution (580)}


\section{Introduction}\label{sec:intro}
A key tracer of galaxy evolution is the change in their chemical composition over time. 
The metallicity of a galaxy is a sensitive observational diagnostic of its past star 
formation history and present-day evolutionary state given that metallicity increases 
with each successive generation of massive star yields 
\citep[e.g.,][]{tosi88, roy95, berg19a, maiolino19}.
Oxygen is an important tracer of metallicity because it is the most abundant element in 
the Universe after H and He and is convenient to observe, with ubiquitous emission lines 
from \ion{H}{2} regions in the rest-frame optical regime. 
While O emission in dwarf and spiral galaxies has been widely observed in the rest-frame  
optical and UV 
\citep[e.g.,][]{kennicutt92, izotov99, vanzee06a, berg12, berg16, senchyna17, berg19a, rogers22}, 
the N emission in these same galaxies has been predominantly traced only in the optical 
through the low-ionization [\ion{N}{2}] \W\W6550,6585 emission lines. 
In general, there is a surprising dearth of detections of the high-ionization N emission 
counterparts in local galaxies, totaling less than 10 galaxies with significant 
detections of either \ion{N}{4}] 
\W\W1483,1486 or \ion{N}{3}] \W1750 \citep[e.g.,][]{mingozzi22,martinez25}.
However, with the advent of JWST, there is a growing prevalence of $z\gtrsim5$ 
galaxies with extreme properties, 
including intense UV N emission
\citep[e.g.,][]{bunker23, isobe23, castellano24, ji24, schaerer24, marques-chaves24,  curti25a, harikane25b, hsiao25, naidu25a, topping25a}.

The first noted, and one of the most distant, examples of extreme rest-frame UV 
N emission comes from the spectroscopically-confirmed $z = 10.6$ galaxy, 
GN-z11.
JWST spectra of GN-z11 revealed surprisingly strong \ion{N}{4}] \W\W1483,1486 and 
\ion{N}{3}] \W1750 emission \citep[e.g.,][]{bunker23} that corresponds to super-solar 
N/O enrichment \citep[$\log({\rm N/O})>-0.25$; e.g.,][]{cameron23}.
Subsequently, enhanced N/O has been reported in a number of high$-z$ galaxies,
including 
GDS 3073 \citep[$z=5.55$;][]{ji24},
RXCJ2248-ID \citep[$z=6.10$;][]{topping24a},
A1703-zd6 \citep[$z=7.04$;][]{topping25a},
CEERS-1019 \citep[$z=8.68$;][]{marques-chaves24},
GNz9p4 \citep[$z=9.38$;][]{schaerer24},
GHZ9 \citep[$z=10.15$;][]{napolitano24},
GHZ2 \citep[$z=12.34$;][]{castellano24}, and
MoM-z14 \citep[$z=14.44$;][]{naidu25a}.
For a review of nitrogen line detections, see \citet{stark25}.
Such strong nebular N$^{+3}$ emission requires a relatively hard ionizing radiation 
field ($\gtrsim47.4$ eV), where models of massive stars predict few photons. 
On the other hand, N$^{+2}$ has a lower ionization potential ($\sim29.6$ eV), 
but statistically significant detections are strikingly rare in integrated galaxy 
spectra \cite[e.g.,][]{berg18,mingozzi22,bunker23,senchyna24} and are only expected to be 
strong at the highest possible nebular temperatures ($\sim2.5\times10^4$ K). 
Furthermore, the timing of the incredibly high N/O abundances reported for the high-
redshift UV N emitters just a few 100 Myr after the Big Bang is unexpected.  

The discovery of significant, rapid nitrogen enhancement so early in the Universe was 
surprising because it contradicts our longstanding understanding of N production. 
In typical chemical evolution modeling, some nitrogen enrichment can occur early 
on via core collapse supernova (CCNe), but substantial nitrogen enrichment only occurs 
100s of Myr after the onset of star formation via asymptotic giant branch (AGB) 
stars \citep[e.g.,][]{vincenzo19,kobayashi20}.
Thus, alternative, faster enrichment methods are needed to explain substantial 
nitrogen enrichment in early galaxies. 
As a result, the necessary ionizing flux and conditions to produce the 
unexpectedly strong N$^{+3}$ and N$^{+2}$ emission observed in galaxies beyond $z\sim5$ 
have been attributed to more extreme sources such as 
active galactic nuclei \citep[AGN;][]{maiolino23},
Wolf Rayet stars \citep[WR; e.g.,][]{senchyna24,watanabe24,gunawardhana25}, 
globular cluster precursors \citep[e.g.,][]{charbonnel23,ji25}, 
super star clusters \citep[e.g.,][]{pascale23},
very massive stars \citep[VMSs: $M_\star > 10^2\ M_\odot$; e.g.,][]{vink2023,shi26}, or
supermassive stars \citep[$M_\star > 10^3\ M_\odot$; e.g.,][]{charbonnel23,nagele23},
tidal disruption events \citep[e.g.,][]{cameron23, watanabe24}, and more. 

Most of our understanding of WR stars has been built from observations of 
individual resolved stars in a handful of galaxies in the Local Group, with almost no
direct spectroscopic evidence for the prevalence of WR stars in more distant galaxies.
To date, only two systems at Cosmic Noon ($z\approx2-3$) have confirmed signatures of 
WR stars: MARTA-4327 at $z=2.224$ \citep[hereafter, M4327;][]{curti25b} and the Sunburst 
Arc at $z=2.37$ \citep[][]{rivera-thorsen24}. 
Extending such detections to earlier cosmic epochs is crucial for understanding the 
role of massive stars in shaping the chemical evolution of galaxies in the first Gyr.

Here we investigate the $z=6.1$ lensed galaxy RXCJ2248-ID3.
RXCJ2248-ID was first identified by \citet{boone13}, \citet{balestra13}, and 
\citet{monna14} and discovered to be a high-ionization, compact, metal-poor,
N-enhanced galaxy by \citet{mainali17}, \citet{schmidt17}, and \citet{topping24a}.
We present extremely deep JWST/NIRSpec observations of RXCJ2248-ID3
that provide the highest-redshift spectroscopic evidence of N-type WR (WN) stars to date,
which provide a physically-consistent mechanism driving its extreme nitrogen 
enrichment \citep{topping24a}.
The remainder of this paper is organized as follows.
The observations and data reduction are briefly described in Section~\ref{sec:obs}, 
followed by a description of the emission line fits, including the broad lines
related to the WR feedback, in Section~\ref{sec:em}.
We present the discovery of WN stars at $z\sim6$ via 
their spectral signatures in Section~\ref{sec:WR}.
We determine new nebular properties and O, C, N, and Si abundances in Section~\ref{sec:abund} 
and compare them to populations of both low- and high-redshift galaxies.
We discuss the source of N enrichment in the early Universe and subsequently estimate 
mass production and timing arguments in Section~\ref{sec:source}.
Finally, we present our conclusions in Section~\ref{sec:conclusions}.
Throughout this work we adopt cosmological parameters of $H_0=70$ km s$^{-1}$ Mpc$^{-1}$, 
$\Omega_{\rm m} = 0.30$, and $\Omega_\Lambda = 0.7$ and the solar abundance pattern from 
\citet{asplund21}.

\section{JWST/NIRSpec Spectra}\label{sec:nirspec}
RXCJ2248 is a galaxy at $z\sim6.1$ that is lensed into multiple images by the 
Abell S1063 cluster ($\alpha=$22:48:44.13,~$\delta=-$44:31:57.50) at a 
redshift of $z =0.348$.
We present an analysis of the brightest image, RXCJ2248-ID3 ($J=25.0$), which has a 
magnification of $\mu\sim7$ \citep{furtak26}.
RXCJ2248-ID was discovered as a $z\sim6$ candidate \citep{boone13,monna14} using 
the 16-band {\it HST} photometry of the CLASH Survey and spectroscopically-confirmed 
via VIsible Multi-Object Spectrograph (VIMOS)/VLT observations by \citet{balestra13}. 
RXCJ2248-ID3 was soon found to be an exciting extreme emission line galaxy via 
ground-based spectroscopy \citep{mainali17}, with strong detections of high-ionization 
emission such as \ion{O}{3}] \W\W1661,1666 and \ion{C}{4} \W\W1548,1550 but no
\ion{He}{2}, suggesting star formation as the ionizing source rather than an AGN.

The early spectra of RXCJ2248-ID3 motivated further rest-UV$+$optical study with {\it JWST}/NIRSpec by 
\citet{topping24a}.
This work performed direct metallicity calculations to show that RXCJ2248-ID3 is 
one of the most extreme N/O-enhanced ($\log(\rm N/O)=-0.39^{+0.11}_{-0.10}$), metal-poor 
($12+\log(\rm O/H)=7.43^{+0.17}_{-0.09}$) galaxies,
with high-ionization ([\ion{O}{3}] \W5008/[\ion{O}{2}] \W3728 = 184) and
high nebular density ($6.4\times10^4 \leq n_e\ (\rm cm^{-3}) \leq 3.1\times10^5$).
They also used spectral energy distribution (SED) fitting with a constant star 
formation history to characterize its low stellar mass ($M_\star\sim10^8\ M_\odot$) 
and young-massive star population ($\sim2$ Myr) of RXCJ2248.
\citet{topping24a}, therefore, suggest that the N/O enrichment may be due to a 
short-lived phase that many $z>6$ bursty galaxies experience. 
In this paper, we build on the work of \citet{topping24a} with new, extraordinarily
deep rest-optical {\it JWST}/NIRSpec observations of RXCJ2248-ID3 from the GLIMPSE-D
Survey, a Director’s Discretionary Time follow-up program described below.

\subsection{Observations and Reduction}\label{sec:obs}
The work presented here uses both the rest-UV {\it JWST}/NIRSpec archival spectra 
from JWST PID 2478 (PI Stark) and new rest-optical {\it JWST}/NIRSpec spectra from the 
GLIMPSE-D Survey, which is an extension of the GLIMPSE Survey.
Properties of RXCJ2248-ID and observation details are presented in Table~\ref{tbl1}.

The GLIMPSE Survey is a large Cycle 2 {\it JWST} program (PID 3293; PIs Atek 
\& Chisholm) that performed ultra-deep NIRCam imaging ($\sim30.8$ mag at $5\sigma$ 
over $0.8 - 5\mu$m) in seven broadband and two medium-band filters of the lensing 
cluster Abell S1063 \citep[][]{atek25}. 
\citet{claeyssens26} performed size and photometric measurements of RXCJ2248-ID 
in the different multiple images. The SED fitting was performed with the Bayesian 
Analysis of Galaxies for Physical Inference and Parameter EStimation 
\citep[\texttt{BAGPIPES};][]{carnall18} code with Binary Population and Spectral Synthesis \citep[\texttt{BPASS} v2.14]{eldridge17} stellar population synthesis burst models and 
\texttt{cloudy} v23.01 photoionization models \citep{chatzikos23, gunasekera23}. 
Priors were used to be physically consistent with the source, i.e., 
high ionization parameter ($-2\leq \log U \leq -1$), low extinction 
($A_v < 0.5$ mag), low-metallicity ($Z < 0.4\ Z_\odot$), and bursty star formation 
($\tau=1$ Myr, i.e., close to a single burst, or $\tau=10$ Myr). 
The resulting best fit has a young age ($t_{\rm age}=1.6^{+11.9}_{-0.9}$ Myr) and
low stellar mass of $M_\star=4.6^{+21.1}_{-2.3}\times10^7\ M_\odot$ but within a 
compact size of $R_{\rm eff} = 19.5^{+13.99}_{-4.72}$ pc such that the stellar mass
surface density is 
$\Sigma_\star=1.1^{+5.2}_{-0.3}\times10^4\ M_\odot{\rm\ pc}^{-2}$.
This value is akin to the highest densities found in globular clusters, similar 
to the ones reported for young star clusters and clumps at high redshift 
\citep{messa2025, claeyssens2025}, and broadly consistent with the conclusions 
presented in \citet{topping24a}. 

\begin{figure}
\begin{center}
	\includegraphics[width=1.0\linewidth,trim=1mm 0mm 0mm 0mm,clip]{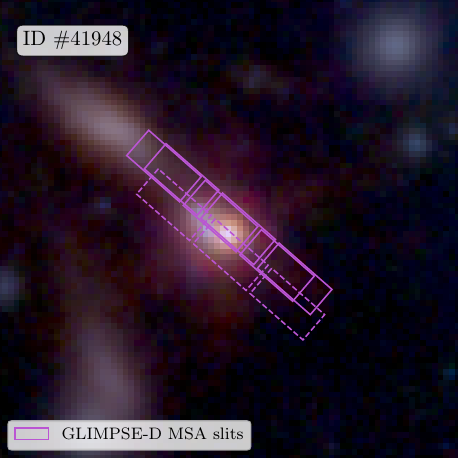}
\caption{
{\it JWST}/NIRSpec MSA slits targeting RXCJ2248-ID3 for each of the three 
exposures in the GLIMPSE-D program.
The two pointings that are closely aligned (solid purple regions) have the
same wavelength coverage, while the pointing offset to the lower left 
(dashed region) is has somewhat reduced blue coverage.
All three pointings were used in the spectrum coaddition. 
\label{fig:msa}}
\end{center}
\end{figure}

Subsequent medium-resolution ($R\sim1000$) spectra of RXCJ2248-ID3 were obtained 
as part of the follow-up GLIMPSE-D Survey: JWST Director’s Discretionary Time 
(DDT) Program 9223 (PIs Fujimoto \& Naidu) targeting a Pop III candidate in 
\citet{fujimoto25} using NIRSpec Multi-Object Spectroscopy (MOS) with the G395M 
grating and F290LP filter.
As part of this program, RXCJ2248-ID3 was observed for a total of 13 exposures using a 3-point nod pattern 
and NRSIRS2 readout, totaling $\sim30$ hours of integration. 
The MSA slit positions covering RXCJ2248-ID3 of the three pointings are shown in 
Figure~\ref{fig:msa}.

We augment the rest-optical GLIMPSE-D data with archival rest-far-UV 
G140M/F100LP and rest-near-UV G235M/F170LP observations from PID 2478 
(PI Stark), covering the rest-frame $\sim1400-4000$ \AA\ range.
This program also observed the G395M/F290LP setting, but we only use 
the significantly deeper GLIMPSE-D G395M observations here. 
Multiple images of RXCJ2248 were identified and observed in program \#2478.
\citet{topping24a} utilized these data by coadding the spectra of the individual images.
In contrast, only the brightest image (ID3) was observed in the GLIMPSE-D Survey.
To ensure consistency, we therefore restricted our analysis to the G140M and G235M 
spectra of ID3 obtained in program \#2478.
As a result, our G140M and G235M measurements are not directly comparable to those 
presented by \citet{topping24a}.

The data were reduced using v0.9.8 of the \texttt{msaexp} pipeline \citep{msaexp}, 
following the standard routines described in 
\citet[][]{deGraaff25,heintz25,valentino25}. 
Briefly, level-2 calibrated products from MAST are subject to a series of custom 
corrections that account for e.g., 1/f noise, bar vignetting, and detector bias. 
We used the ``local'' nodded background subtraction. 
2D spectra were drizzled onto a common wavelength grid and 1D spectra were optimally 
extracted using a profile model that accounts for e.g., the wavelength-dependent PSF 
and offsets from the nominal position expected from the catalog.
Line centers were measured for the strongest emission lines in the G395M spectrum 
(i.e., H$\delta$, H$\gamma$, [\ion{O}{3}] \W4364, H$\beta$, [\ion{O}{3}] \W\W4960,5008, 
\ion{He}{1} \W5877, H$\alpha$, \ion{He}{1} \W7067) and used to determine a redshift of 
$z=6.1025\pm0.0013$.
Note that the bluest portion of the G395M tends to favor a slightly lower redshift 
(i.e., $z\sim6.1000$), while the reddest portion favors a slightly higher redshift 
(i.e., $z\sim6.1034$).
The three individual 1D extracted spectra were then normalized to the common 
continuum flux scale of the first spectrum at rest-wavelengths of 
$\sim6000-62000$ \AA\ prior to coadding.
Spectral coaddition was performed as a weighted average using the inverse 
variance as the weight.

\begin{figure*}
\begin{center}
	\includegraphics[width=1.0\linewidth,trim=0mm 0mm 0mm 0mm,clip]{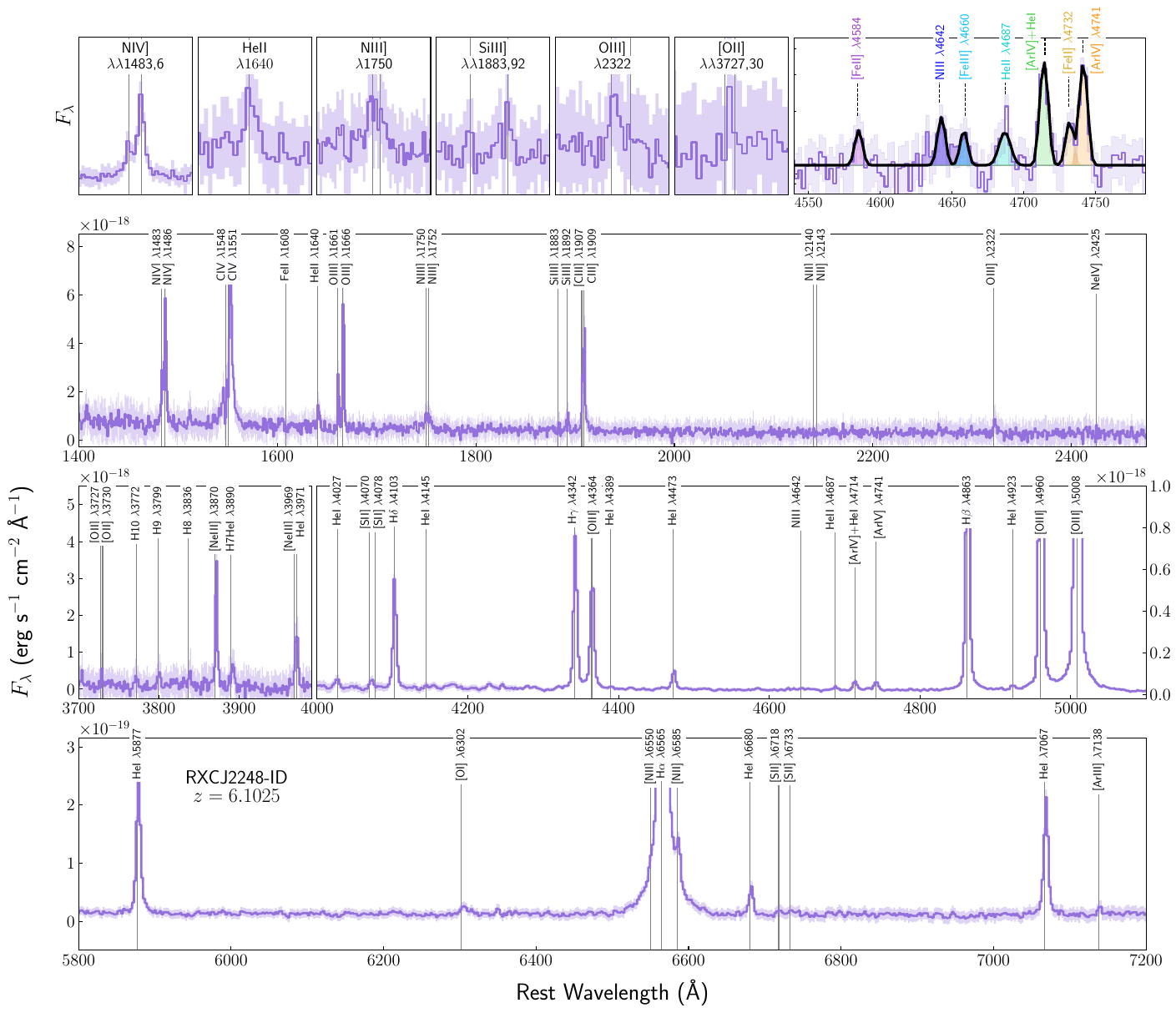}
\caption{
{\it JWST}/NIRSpec rest-frame UV and optical spectra of RXCJ2248-ID3 
highlighting the first object known with simultaneously detected emission from
N$^+$, N$^{+2}$, and N$^{+3}$ \citep[see, also,][]{topping24a} and WR features.
The 2nd row shows the main emission UV emission line detections from the archival 
G140M/F100LP spectrum, with significant detections of several high-ionization 
emission lines, including \ion{N}{4}] \W\W1483,1486, \ion{C}{4} \W\W1548,1550, 
\ion{He}{2} \W1640, \ion{O}{3}] \W\W1661,1666, \ion{N}{3}] \W1750, and \ion{C}{3}] 
\W\W1907,1909.
The 3rd row shows the blue end of the optical spectrum, where the left hand panel 
shows the archival G235M/F170LP spectrum, which includes the low-ionization 
[\ion{O}{2}] \W\W3727,3730 doublet.
The right hand panel of the 3rd row and the 4th row shows the extremely high 
S/N GLIMPSE-D optical spectrum, enabling detections of several weak features.
Note that 
Some of the important features to this work are highlighted in the zoom in panels
in the top row. 
In particular, the last panel reveals the most distant WR detection to date, 
with the \W4650 WR bump showing emission from \ion{N}{3} \W4642, indicative 
of nitrogen enrichment from WN stars. 
Note that not all line labeled correspond to line detections. 
\label{fig:spec}}
\end{center}
\end{figure*}

The resulting spectrum, shown in Figure~\ref{fig:spec}, covers an observed 
wavelength range $\sim2.8-5.5\ \mu$m, which corresponds to a rest-optical 
range of $\sim3900-7740$ \AA.
Note that the third pointing (dashed slits in Figure~\ref{fig:msa}) has reduced 
wavelength coverage such that the blue end begins at $\sim4265$.
The deep GLIMPSE-D spectra provide unparalleled signal-to-noise (S/N; $>5$ at 5100 \AA\ continuum)
that enable rest-optical diagnostics typically reserved for nearby galaxies.


\begin{deluxetable}{rcc}
\setlength{\tabcolsep}{7pt}
\tablecaption{Properties of RXCJ2248-ID3}\label{tbl1}
\tablehead{\multicolumn{3}{c}{JWST/NIRSpec Observations} }
\startdata
\CH{Grating/Filter} & \CH{$t_{\rm exp}$ (s)}  & \CH{PI/PID} \\
\hline
G140M/F100LP        & 6215                    & Stark/2478  \\
G235M/F170LP        & 1576                    & Stark/2478  \\
G395M/F290LP        & 107228                  & Fujimoto \& Naidu/9223 \\ 
\hline\hline
\multicolumn{3}{c}{Measured Properties} \\
\hline
\CH{Property}           & \CH{Value}    & \CH{Reference} \\
\hline
R.A.                    & $+$22:48:45.81            & This work          \\
Decl.                   & $-$44:32:14.95            & This work          \\
$z$                     & $6.1025\pm0.0013$         & This work          \\
$\mu$                   & 6.8877                    & \citet{furtak26}   \\
$R_{\rm eff}$ (pc)      & $19.5^{+13.99}_{-4.72}$   & \citet{claeyssens26} \\
$M_\star$ ($M_\odot$)   & $4.6^{+21.1}_{-2.3}\times10^7$   & \citet{claeyssens26} \\
$\Sigma_\star$ ($M_\odot$ pc$^{-2}$)
                        & $1.1^{+5.2}_{-0.3}\times10^4$   & \citet{claeyssens26} \\
SFR$_{\rm H\alpha}$ ($M_\odot$ yr$^{-1}$)
                        & \SFR                     & This work, \S~\ref{sec:Nmass} \\
SFR$_{\rm SED,1\ Myr}$ ($M_\odot$ yr$^{-1}$) 
                        & 4.7                      & \citet{claeyssens26} \\
SFR$_{\rm SED,10\ Myr}$ ($M_\odot$ yr$^{-1}$)
                        & 4.1                      & \citet{claeyssens26}\\
$\Sigma_{\rm SFR}$ ($M_\odot$ yr $^{-1}$ kpc$^{-2}$)   
                        & $1.34\times10^3$& This work \\
$t_{\rm age}$ (Myr)     & $1.6^{+11.9}_{-0.9}$      & \citet{claeyssens26} \\
12+log(O/H)             & $\loh\pm\lohe$            & This work, \S~\ref{sec:OH} \\
log(N/O)                & $\lnoave\pm\lnoavee$      & This work, \S~\ref{sec:NO}
\enddata
\tablecomments{
{\it Top:} JWST/NIRSpec observations of RXCJ2248-ID3, including archival 
observations from PID 2478 (PI: Stark) and very deep GLIMPSE-D observations 
from PID 9223 (PI: Fujimoto \& Naidu).
Columns 1-3 list the grating/filter, exposure time, and 
principle investigator/PID.
{\it Bottom:} Measured global properties of RXCJ2248-ID3.
The R.A. and decl. are the extraction coordinates for RXCJ2248-ID3.
The redshift was determined from the GLIMPSE-D spectrum emission lines.
GLIMPSE imaging was used to determine the lensing model magnification, 
$\mu$.
Effective radius of the RXCJ2248-ID3 clump, stellar mass, and current massive 
star population age are from the SED modeling of \citet{claeyssens26}, while the SFR 
was determined from both the SED fitting and  the narrow-component, collisions-corrected H$\alpha$ 
flux (see \S~\ref{sec:Nmass}), all corrected for the lensing factor.
The star formation rate surface density was determined using the SFR$_{\rm H\alpha}$.
The metallicity and relative N/O abundance were determined 
using the direct method. }
\end{deluxetable}


\subsection{Emission Line Measurements}\label{sec:em}
In order to perform a consistent analysis of our data, we measure 
emission line fluxes for both the archival spectra and the new 
GLIMPSE-D spectra presented here.
We fit neighboring emission lines simultaneously using Gaussian 
profiles with the \texttt{lmfit} package \citep{newville15} in \texttt{python}.
Purely nebular lines (i.e., lines without possible stellar contributions 
or resonant effects) close in wavelengths were constrained to have 
the same full width at half-maximum (FWHM) velocity widths.
Additionally, the relative wavelength spacing between lines was 
constrained to  laboratory values and doublets with constant flux 
ratios set by atomic physics were constrained to their theoretical 
values, with small uncertainty allowances.
The uncertainties on the line fluxes were estimated as the standard 
error derived from the least-squares minimization in \texttt{lmfit}, 
which considers the uncertainty on the Gaussian profile and linear 
continuum.

Broad emission components are clearly visible at the base of some of 
the emission lines in the GLIMPSE-D spectrum of RXCJ2248-ID3.
Such broad emission features can be produced by stellar winds, 
shocks, or turbulence. 
Since \ion{He}{2} \W1640 and \W4687 emission lines can be affected 
by stellar winds, we fit these features with an unconstrained 
Gaussian width.
Using the \texttt{jwst-msa} package \citep{deGraaff24}, we deconvolved
all measured FWHMs with the modeled wavelength-dependent line spread 
function (LSF). 
We found the \ion{He}{2} lines to be broadened compared to purely 
nebular lines.
For the \ion{He}{2} \W4687 line, the velocity width is $528\pm100$ 
km s$^{-1}$, which is more than two times broader than the narrow
nebular H$\beta$ component with $v_{\rm FWHM}=243\pm25$ km s$^{-1}$.\looseness=-4

The strongest rest-optical H (H$\gamma$, H$\beta$, and H$\alpha$) and [\ion{O}{3}] 
(\W4364, \W\W4960,5008) emission lines have complex profiles with both narrow and 
broad emission components.
Such broad components may also be present in the rest-UV and  fainter rest-optical 
emission lines, but none are obvious given the lower S/N of these emission features 
and/or underlying continuum. 
To fit these profiles, we tested three different multicomponent 
profile combination fits for the H$\alpha +$[\ion{N}{2}] complex.
For all three fits, the narrow H$\alpha$ and [\ion{N}{2}] \W\W6550,6585 
lines were fit by Gaussians with a single velocity width, but the 
broad component was fit with either:
(1) a single Gaussian profile, (2) two Gaussian profiles, or 
(3) a single exponential profile. 
The single broad Gaussian profile fit had strong residuals near the 
center of the broad component, so did not provide a good fit to the 
observed emission profile.
Both the double Gaussian profile and the exponential profile provided 
relatively good visual fits, but the double Gaussian fit had a lower reduced 
chi-squared ($\chi^2_{\rm 2Gauss}=0.93$ versus $\chi^2_{\rm exp.}=1.52$) and 
Bayesian inference criteria (BIC$_{\rm 2Gauss}=36$ versus (BIC$_{\rm exp.}=87$), 
and so was adopted as the better statistical fit.\looseness=-2

\begin{figure*}
\begin{center}
	\includegraphics[width=1.0\linewidth, trim= 0mm 0mm 0mm 0, clip]{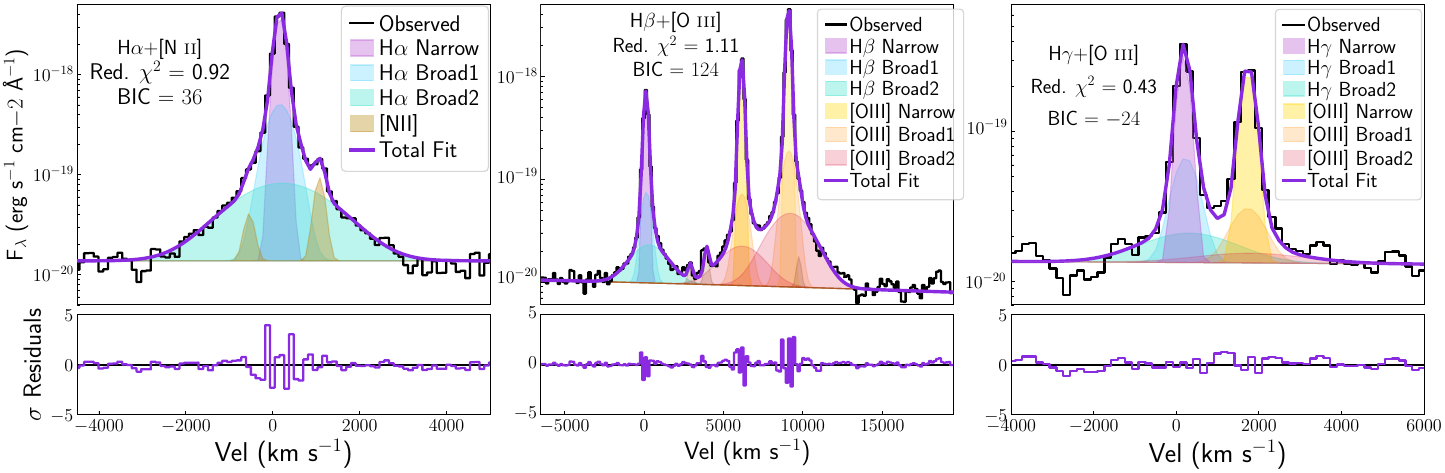}

\caption{
multicomponent emission line fits to the GLIMPSE spectrum of RXCJ2248-ID for 
H$\alpha$ \W6565 $+$ [\ion{N}{2}] \W\W6549,6585 (left panels), 
H$\beta$ \W4863 $+$ [\ion{O}{3}] \W\W4960,5008 (middle panels), and
H$\gamma$ \W4342 $+$ [\ion{O}{3}] \W4364 (right panels).
When fit with single, narrow Gaussian components (e.g., purple and yellow
filled Gaussians), all three line complexes show strong, broad component
residual flux. 
The resulting best fit to each line is comprised of a single narrow 
Gaussian plus two broad Gaussians, where the relevant component velocity 
widths are tied together:
The H$\alpha$ \W6565 $+$ [\ion{N}{2}] \W\W6549,6585 complex fit provided
the velocity width constraints for the H Balmer line narrow (purple Gaussians) 
and broad components (blue and green Gaussians) and, subsequently, the
H$\beta$ \W4863 $+$ [\ion{O}{3}] \W\W4960,5008 fit constrained the [\ion{O}{3}]
narrow (yellow Gaussian) and broad (orange and red Gaussians) velocity
widths that were then used in the H$\gamma$ \W4342 $+$ [\ion{O}{3}] \W4364 fit. 
Note that additional faint lines (e.g., \ion{He}{1} \W5017) were included in 
the fit in the middle panel.
Careful accounting for the residual broad flux has significant impact 
on the derived nebular reddening, temperature, metallicity, and N/O abundance.
\label{fig:fits}}
\end{center}
\end{figure*}

The right panel of Figure~\ref{fig:fits} shows the best multicomponent 
fit to the H$\alpha +$[\ion{N}{2}] complex.
Since all kinematically-similar lines in the Balmer emission series 
arise from the same gas, we expect the H$\beta$ and H$\gamma$ profiles to 
be well fit by scaling the H$\alpha$ best fit. 
Therefore, we constrained the velocity widths of the H$\beta$ and 
H$\gamma$ emission components to match the narrow $+$ double broad 
Gaussian H$\alpha$ fit, accounting for the wavelength-dependent LSF.
We found excellent fit results, with similarly 
small reduced-$\chi^2$ and BIC values.
This means that the \ion{H}{1} lines are well fit by a profile with 
(1) a strong, narrow ($\sim$250 km s$^{-1}$) nebular component, 
(2) a moderate ($\sim20$\%\ of total flux), broad component 
($\sim$670 km s$^{-1}$), and 
(3) a weak ($\sim10$\%\ of total flux), very broad 
($\sim$2530 km s$^{-1}$) component.

The [\ion{O}{3}] \W\W4960,5008 doublet lines are also well fit by a narrow 
Gaussian plus double Gaussian broad component profile, with the relative 
fluxes of each component constrained to the theoretical ratio.
While the narrow component FWHM was set to the velocity width of the narrow 
Balmer lines, convolved with the LSF, we allowed the FWHM of the two broad 
[\ion{O}{3}] components to vary freely and found widths of 
$\sim890$ km s$^{-1}$ and $\sim2980$ km s$^{-1}$, respectively. 
The similarity between the [\ion{O}{3}] and \ion{H}{1} velocity widths 
of the broad components argues against emission from an AGN directly 
(where high densities cause collisional de-excitation of [\ion{O}{3}]) 
and is more consistent with stellar or AGN driven winds
\citep[e.g.,][]{izotov08,grafener15,grafener17,burke21}.
Interestingly, the broad components of the \ion{H}{1} lines compose 
a larger fraction of their total flux ($\sim20$\%\ and 10\%, respectively) 
than [\ion{O}{3}] ($\sim$10\%\ and 5\%, respectively).

The resulting fit to the H$\beta+$[\ion{O}{3}] \W\W4960,5008 complex is 
shown in the middle panel of Figure~\ref{fig:fits} to be an excellent fit, 
with minimal residuals.
The exquisite S/N of the GLIMPSE spectrum also reveals 
broad wings on the [\ion{O}{3}] \W4364 profile, as seen in the left 
panel of Figure~\ref{fig:fits}.
Therefore, we also applied the narrow Gaussian plus double Gaussian broad 
component profile to [\ion{O}{3}] \W4364, constraining the velocity widths 
to the values measured for [\ion{O}{3}] \W\W4960,5008.

Double broad components with similar velocity widths (750 and 2500 
km s$^{-1}$, respectively) are seen in the $z\sim0$ extreme emission 
line galaxies, J1044+0353 and J1418+2102, reported in \citet{berg21}.
However, each broad component observed in these nearby analogs only 
accounts for 1\%–3\%\ of the total \ion{H}{1} flux. 
This sort of broad component emission from the Balmer H and [\ion{O}{3}] 
lines with widths ($1000-2000$ km s$^{-1}$) and fractional fluxes of 
1\%–2\%\ is commonly found in spectra of blue compact dwarf galaxies 
\citep[BCDs; e.g.,][]{izotov06b,izotov07}.
This suggests that bulk motion of the gas is typical in these metal-poor, 
bursty environments, but for a larger mass of gas in RXCJ2248-ID3.

The sensitive accounting of broad component emission afforded by the deep 
GLIMPSE-D spectra is important, because even a small fraction of broad emission 
around H emission line can significantly affect the fit to weak lines 
such as [\ion{N}{2}] \W\W6550,6585 \citep[e.g.,][]{berg21}. 
In RXCJ2248-ID, the broad components compose a significant fraction of the 
total H and [\ion{O}{3}] fluxes, and so are critical to properly measure 
not only the [\ion{N}{2}] \W6585 emission, but also the [\ion{O}{3}] \W4364, 
H$\beta$, [\ion{O}{3}] \W\W4960,5008, and H$\alpha$ narrow line fluxes.
For this reason we adopt the narrow-line fluxes from our best multicomponent 
fits for the remaining analysis; we reserve further investigation of 
the the broad emission for a forthcoming paper.\looseness=-2

As noted above, the UV spectra do not have sufficient S/N to 
decompose narrow and possible broad components. 
As a result, density diagnostics and relative abundance ratios determined 
from UV line ratios may include contributions from multiple kinematic components. 
If the broad components arise from gas with distinct physical conditions, 
this could introduce systematic offsets. 
We test the level of bias possible due to broad 
component contamination of narrow-line fluxes by adopting the relative 
narrow and broad component profiles of [\ion{O}{3}] \W5008 as a template for 
collisionally-excited lines. 
The broad component areas overlap with the narrow profile such that the 
broad components are responsible for 8.6\%\ and 2.2\%\ of the narrow 
component flux, or 10.8\%\ in total. 
We use this fraction to set the upper contamination limit of potential
broad components to the UV emission lines and determine the impact on 
nebular density, temperature, and abundance calculations in Section~\ref{sec:broad_abund}. 
\looseness=-2

\subsection{Reddening Correction}\label{sec:red}
The observed Balmer decrement of the narrow H$\alpha$/H$\beta$ lines is 
$F_{\rm H\alpha}/F_{\rm H\beta}=\FHAHB$, implying either a moderate amount of 
dust is present or collisional-enhancement of H$\alpha$. 
This value disagrees with the results of \citet[][]{topping24a}, who measured 
an observed decrement of $2.55\pm0.05$ that they found to be consistent with 
no dust attenuation. 
Similarly, \citet{crespo-gomez25} used high-resolution NIRSpec/G395H data to fit 
multiple component Balmer decrements for RXCJ2248-ID3, finding a narrow-component
$F_{\rm H\alpha}/F_{\rm H\beta}=2.7$ that is consistent with no attenuation, but
broad and very-broad component decrements of 4.3 and 6.6, respectively, that 
imply differential extinction.
We too find higher $F_{\rm H\alpha}/F_{\rm H\beta}$ ratios for the broad components,
but the source of this increase is not clear;
it could indicate higher dust in the broad component gas, as suggested by 
\cite{crespo-gomez25}, or result from significant collisional enhancement 
of H$\alpha$.

Fortunately, the GLIMPSE-D spectrum provides a significant increase in S/N in the 
continuum, allowing for more robust fitting of  broad components, including in the 
H$\gamma$ and [\ion{O}{3}] \W4364 and \W5008 lines.
Fitting the broad components directly in the [\ion{O}{3}] lines offers the advantage over 
previous works that we do not need to correct for broad component contamination with 
differential extinction in our $T_e$ calculation. 
Furthermore, by fitting the broad components in H$\gamma$ we were able to examine of 
the narrow component H$\beta$/H$\gamma$ ratio, finding a decrement of 
$F_{\rm H\beta}/F_{\rm H\gamma}=\FHBHG$ that is consistent with very little dust.
Note that we do not consider the H$\beta$/H$\delta$ ratio here as the H$\delta$
line is not strong enough to robustly fit the broad components in a consistent
manner with the profile fitting of the H$\gamma$, H$\beta$, and H$\alpha$ lines.

The reddening due to dust, characterized by $E(B-V)$, was determined by 
comparing the observed Balmer decrements with the theoretical Balmer ratios 
assuming case B and an extinction law, for which we tested the parameterization 
from both \cite{cardelli89} and \cite{calzetti00}.
The $E(B-V)$ value for a given Balmer ratio was determined iteratively until 
convergence, recomputing the \ion{H}{1} theoretical ratio 
using the updated electron temperature from the reddening-corrected [\ion{O}{3}] 
\W4364/\W5008 flux ratio and density from the reddening-corrected \ion{N}{4}] 
\W1483/\W1487 flux ratio in each iteration.
In this way, the reddening, electron temperature, and electron density were 
solved for simultaneously and consistently. \looseness=-2

A greater enhancement of the observed $F_{\mathrm{H}\alpha}/F_{\mathrm{H}\beta}$ decrement than 
of the $F_{\mathrm{H}\beta}/F_{\mathrm{H}\gamma}$ decrement can arise under high-density 
conditions, where collisional excitation selectively enhances the lowest excited level ($n=2$; requires lowest energy to excite), leading to higher H$\alpha$ 
flux relative to H$\beta$ and H$\gamma$.
To assess whether such an enhancement is physically plausible, we examined the 
\texttt{Cloudy} photoionization models \citep{chatzikos23, gunasekera23} presented 
in \citet{martinez25}, which span a wide range of nebular densities 
(up to $n_e=10^9$ cm$^{-3}$).
For the nebular conditions determined in this work (i.e., $T_e$, $\log U$, $Z$, N/O; 
see Section~\ref{sec:neb_prop}), densities of $n_e\sim10^6$ cm$^{-3}$ are needed 
to produce the observed H$\alpha$ enhancement while minimally affecting H$\beta$ 
and H$\gamma$.
Although this density is roughly an order of magnitude higher than the values 
measured in \citet{topping24a} and in this study (Section~\ref{sec:temden}; 
Table~\ref{tbl3}), it could indicate that the ISM contains unresolved clumps 
of even higher density than the volume-weighted values probed by the density
diagnostics used in this work. 
We, therefore, attribute the observed H$\alpha$ excess to collisional enhancement.

Accordingly, we adopted the reddening derived from H$\gamma$/H$\beta$, 
$E(B-V)=\ebv^{+\ebveu}_{-\ebved}$ mag using \citeauthor{cardelli89}
(the \citeauthor{calzetti00} value is similar at $E(B-V)=0.050\pm0.1215$ mag), 
and corrected all emission lines for the resulting (minimal) dust attenuation.
We used the \citet{calzetti00} reddening law for the rest-UV emission lines 
($\lambda<3200$ \AA) and the \citet{cardelli89} reddening law for the 
rest-optical emission lines ($\lambda>3200$ \AA). 
After applying the reddening correction, the H$\alpha$/H$\beta$ ratio still shows 
a collisional excess of \HAexc\ above the theoretical value; we correct for this 
excess and report a final 
$F_{\rm H\alpha}=$\FHaCorr\ erg s$^{-1}$ cm$^{-2}$.

The adopted reddening and dereddened line intensities are listed in Table~\ref{tbl2}
for all line fluxes used in this work.
Note that rest-UV and rest-optical lines should not be compared or combined in line ratios. 
As the rest-UV and rest-optical spectra were obtained during different observing 
runs with distinct pointings and strategies, we report the UV lines relative to 
$F_{\rm CIII] \lambda\lambda1907,1909}\times100$ and the optical lines relative 
to $F_{\rm H\beta}\times100$, without applying any relative scalings between 
the two datasets.


\begin{deluxetable}{lcc}
\setlength{\tabcolsep}{15pt}
\tablecaption{Rest UV$+$Optical Emission Line Fluxes}
\tablehead{
\CH{Ion$+$Wavelength (\AA)} & \CH{$I$(\W)/$I$(\ion{C}{3}])} & \CH{EW (\AA)}}
\startdata
\ion{N}{4}] \W1483.33       & 42.78$\pm$1.61    & 6.67 \\
\ion{N}{4}] \W1486.50       & 102.0$\pm$0.82    & 15.9 \\
\ion{He}{2} \W1640.42       & 22.46$\pm$1,97    & 4.88 \\
\ion{O}{3}] \W1666.15       & 85.84$\pm$0.59    & 18.9 \\
\ion{N}{3}] \W1750$^a$      & 38.59$\pm$0.64    & 9.18 \\
\ion{Si}{3}] \W1883.00      & 5.01$\pm$3.25     & 1.32 \\
\ion{Si}{3}] \W1892.03      & 8.25$\pm$1.98     & 2.21 \\
\ion{C}{3}] \W1906.68       & 35.11$\pm$0.31    & 9.65 \\  
{[\ion{C}{3}]} \W1908.73    & 64.89$\pm$0.25    & 17.9 \\
[2ex]
\hline
{Ion$+$Wavelength (\AA)}    & {$I$(\W)/$I$(H$\beta$)} \\
\hline
{[\ion{O}{2}]} \W3728$^a$   & 4.09$\pm$2.05    & 6.22 \\
H$\gamma$ \W4341.66$^b$     & 47.41$\pm$3.07   & 73.8 \\
{[\ion{O}{3}]} \W4364.44$^b$& 42.45$\pm$1.92   & 66.5 \\
\ion{He}{1} \W4472.73       & 8.90$\pm$0.39    & 27.8 \\
\ion{N}{3} \W4641.94        & 1.40$\pm$0.20    & 4.4  \\
\ion{He}{2} \W4687.01       & 1.33$\pm$0.29    & 4.2  \\
{[\ion{Ar}{4}]} \W4712.69$^c$&2.30$\pm$0.27    & 10.3 \\ 
\ion{He}{1} \W4714.46$^c$   & 1.91$\pm$0.19    & 3.0  \\
{[\ion{Ar}{4}]} \W4741.49   & 4.10$\pm$0.26    & 13.0 \\
H$\beta^c$ \W4862.71        & 100.0$\pm$4.4    & 356  \\
{[\ion{O}{3}]} \W4960.29$^b$& 230.5$\pm$9.0    & 877  \\ 
{[\ion{O}{3}]} \W5008.24$^b$& 708.9$\pm$27.5   & 2791 \\
H$\alpha$ \W6564.60$^{b,d}$ & 331.8$\pm$14.4   & 1755 \\
H$\alpha$ \W6564.60$^{b,e}$ & 274.1$\pm$11.9   & 1457 \\
{[\ion{N}{2}]} \W6585.27    & 7.08$\pm$0.91    & 12.8 \\
{[\ion{S}{2}]} \W6718.29    & 0.68$\pm$0.29    & 4.78 \\
{[\ion{S}{2}]} \W6732.67    & 0.81$\pm$0.30    & 4.74 \\
\hline
$E(B-V)$                    & $\ebv^{+\ebveu}_{-\ebved}$  \\
$F_{\rm CIII]}$             &11.58$\pm$0.49 \\
$F_{{\rm H}\beta}^b$        &6.94$\pm$0.15 
\enddata
\tablecomments{
Reddening-corrected emission-line intensities of lines used in this analysis 
from the archival rest-UV and GLIMPSE rest-optical {\it JWST}/NIRSpec spectra 
for RXCJ2248-ID3. 
Note that no scaling was performed between the archival UV and GLIMPSE-D 
optical pointings (not needed for this work).
Thus, UV fluxes are given relative to the $F_{\rm CIII]\lambda\lambda1907,09}\times100$
and optical fluxes are given relative to $F_{\rm H\beta}\times100$.
The last three rows list the dust attenuation derived using the \citet{cardelli89}
reddening law and the rest-frame \ion{C}{3}] \W\W1907,09 and H$\beta$ flux in units of
$10^{-18}$ erg s$^{-1}$ cm$^{-2}$. 
Additionally, the fluxes reported here are for a single image of RXCJ2248 (ID3), 
whereas \citet{topping24a} report fluxes for coadded spectra of multiple images,
and so correct for the lensing.\\
$^a$Note that \ion{N}{3}] \W1750 and [\ion{O}{2}] \W3728 fluxes are the 
integrated values for the \ion{N}{3}] \W\W1746,1748,1749,1752,1754 quintuplet 
and [\ion{O}{2}] \W\W3727,3730 doublet, respectively. \\
$^b$Emission line profile was best fit with a narrow Gaussian $+$ two broad
Gaussian components; only the corrected narrow line flux is listed here
(see \S~\ref{sec:em} and Figure~\ref{fig:fits}). \\
$^c$ [\ion{Ar}{4}] \W4713$+$\ion{He}{1} \W4714 is a blended line 
profile at the observed resolution. 
Thus, the [\ion{Ar}{4}] \W4713 is determined by subtracting the
\ion{He}{1} \W4714 flux, which is predicted from the \ion{He}{1} 
\W4473 flux. \\
$^d$Uncorrected for collisional-excitation. \\
$^e$Corrected for collisional-excitation. }\label{tbl2}
\end{deluxetable}

\section{Wolf Rayet Stars at \MakeLowercase{\it z} = 6.1}\label{sec:WR}
The WR stage of massive star evolution is an important, short-lived
phase that can have significant effects on the chemical composition of the 
local ISM.
We provide a brief overview here \citep[see, e.g.,][for a more thorough review]{crowther07}.
WR stars are massive stars that have entered the core He-burning phase and 
have lost their outer envelope either via strong stellar winds or due to 
binarity effects (i.e., stripping via Roche Lobe overflow or mergers).
The first phase of WR stars occurs when the outer H layer has been ejected 
revealing the H core-burning products such that their spectra are 
characteristically He and N rich, but are H-poor.
Such stars are known as nitrogen-type WR, or WN, stars, and are often 
identified by strong \ion{N}{3}, \ion{N}{4}, and \ion{N}{5} emission lines, 
especially the broad optical ``blue bump'' near \W4650. 
The  blue bump is a complex of features, including \ion{N}{3} \W\W4634,4642, 
\ion{C}{3} \W4649,4667, \ion{Fe}{3} \W4660, and \ion{He}{2} \W4687.
Subsequently, stars that are massive enough for core He-burning and for 
their winds to remove their outer He envelope and expose the produced C, 
enter the WR-carbon, or WC, phase.
WC stars also have strong, broad \ion{He}{2} emission and strong C and 
O emission such that they are identified by the optical WR \ion{C}{4} 
\W\W5803,5814 doublet (the ``red bump''). 
As a result, the typically very strong winds of the WR phase can produce 
significant N enrichment during the WN phase and drive strong C ejection 
during the WC phase.
After the WC phase, a WR-oxygen phase may ensue, but we forgo discussion of this phase here. 

The rest-frame UV and optical spectra shown in Figure~\ref{fig:spec} can be used to characterize the WR nature of the
stellar population in RXCJ2248-ID3.
Both the UV and optical \ion{He}{2} emission features are kinematically broadened 
compared to the narrow nebular emission features in RXCJ2248-ID3, indicative of 
WR or VMS winds. 
\citet{martins23,martins25} have shown that young star forming regions 
dominated by VMSs can be distinguished from WR stars using the morphology of the 
blue and red bumps. 
In particular, VMSs produce blue bumps with \ion{He}{2} \W4687 emission 
but little to no \ion{N}{3} emission and red bumps with narrow \ion{C}{4} 
\W\W5803,5814 emission.
Thus, strong detections of \ion{N}{3} in the blue bump favor a WN interpretation 
\citep[e.g.,][]{martins23,berg24,rivera-thorsen24}.

The upper right hand panel of Figure~\ref{fig:spec} highlights the blue bump 
spectral regime, showing weak, broad \ion{He}{2} and \ion{N}{3} \W4642 in 
RXCJ2248-ID3, both of which are characteristic of metal-poor WN stars.
Just redward of the \ion{N}{3} \W4642 line in the blue bump (but blueward of 
[\ion{Fe}{3}]), a second less prominent emission feature is seen, but it is 
difficult to determine whether this is due to \ion{C}{3} or \ion{O}{2} emission, 
or both.
Furthermore, the red \ion{C}{4} bump isn't detected, suggesting little to no 
contributions from WC stars or VMSs in the spectrum. 
Thus, we only significantly detect the blue WR bump, suggesting that WN stars 
are likely present.

\section{Nebular Properties}\label{sec:neb_prop}
Using the updated narrow component emission line fits presented in 
Section~\ref{sec:em}, we determined the nebular properties of RXCJ2248-ID3.
Following \citet{berg21}, we adopt the four-zone ionization model to 
account for the high ionization emission observed.
In this model, the ionization potential energy ranges of N$^{+}$, S$^{+2}$, 
O$^{+2}$, and He$^{+2}$ define the low-, intermediate-, high-, and very 
high-ionization zones, respectively.
For all calculations, we use the \texttt{PyNeb} package in \texttt{python} with the 
atomic data adopted in \citet{berg19a}, which includes a six-level atom model for 
oxygen in order to utilize the UV \ion{O}{3}] \W1666 line.
Below we determine temperatures and densities, although $T_e$(O$^{+2}$) and 
$n_e$(N$^{+3}$) were co-determined during the iterative reddening calculation (see 
\S~\ref{sec:red}) in \S~\ref{sec:temden}, ionization parameters in \S~\ref{sec:logu},
and abundances in \S~\ref{sec:abund}.

We note that the UV spectra do not have sufficient S/N to decompose narrow 
and broad components following the same method as the optical lines. 
As a result, density diagnostics and abundances determined from UV lines may include 
contributions from multiple kinematic components, while optical temperatures, densities, 
and abundances are derived from narrow components alone. 
If the broad component arises from gas with distinct physical conditions, 
this could introduce systematic offsets.
For this reason, we examine the potential impact of UV broad components in 
\S~\ref{sec:broad_abund}.

\subsection{Temperature and Density}\label{sec:temden}
One of the unique characteristics of RXCJ2248-ID3 is its large 
number of density-sensitive emission line ratios.
\citet{topping24a} previously reported densities from the three UV 
line ratios of \ion{Si}{3}] \W1883/\W1892, characterizing the 
intermediate-ionization zone, \ion{C}{3}] \W1907/\W1909, characterizing 
the intermediate- to high-ionization zone, and \ion{N}{4}] \W1483/\W1486, 
characterizing the high- to very high-ionization zone.
The new high-S/N optical spectra enables us to measure, for the first time, 
densities from the low-ionization [\ion{S}{2}] \W6717/6731 ratio and the
high- to very high-ionization [\ion{Ar}{4}] \W4713/\W4741 ratio.

We use our narrow component de-reddened flux measurements to compute densities 
for all five line ratios and the high-ionization zone temperature 
from the [\ion{O}{3}] \W4364/\W5008 ratio.
The high-ionization zone $T_e$(O$^{+2}$) and $n_e$(N$^{+3}$) were simultaneously 
determined during the iterative reddening calculation in \S~\ref{sec:red} to account 
for the sensitivities of both diagnostics.
If the low density limit was assumed instead ($n_e\lesssim10^2\ {\rm cm}^{-3}$), 
as is common practice at low-redshift, the observed [\ion{O}{3}] \W4364/\W5008 flux 
ratio would lead to unphysical temperatures (i.e., above the limit set by H cooling 
of $\sim2.5\times10^4$ K).
Thus, a physical and robust solution requires high densities to properly 
account for the reduced \W5008 flux due to collisionally de-excitation.
Furthermore, \citet{martinez25} recently showed that densities derived from both 
optical and UV diagnostics under-predict the true volume-averaged density in 
multiphase, high-density systems, with more severe under-prediction from the 
optical diagnostics.
Therefore, it is necessary to use UV density diagnostics in high-density 
environments, though the true density will still be underestimated in 
multiphase gas \citep[see, e.g., Figure 11 of][]{martinez25}.

For the high ionization zone, we found a $T_e$(O$^{+2}$)$=\thigh\pm\thighe\times10^4$ 
K and $n_e$(N$^{+3}$)$=\nNIV^{+\nNIVeu}_{-\nNIVed}\times10^5$ cm$^{-3}$, 
which is consistent with the density of 
$n_e$(N$^{+3}$)$=3.1^{+0.5}_{-0.4}\times10^5$ cm$^{-3}$ reported by \citet{topping24a},
but lower than their temperature of $2.46\pm0.26\times10^4$ K
due to our broad component fits of both [\ion{O}{3}] \W4364 and \W5008.
Adopting our $T_e$(O$^{+2}$) as the high-ionization temperature 
($T_{e,{\rm high}}$), we then applied the $T_e-T_e$ relations of \citet{garnett92} to 
estimate the intermediate-ionization temperature ($T_{e,{\rm int.}}$) and low-ionization 
temperature ($T_{e,{\rm low}}$).\looseness=-2

The determined temperatures were used for the subsequent density 
calculations in their respective ionization zones. 
Note that the [\ion{Ar}{4}] \W4713 and \ion{He}{1} \W4714 
lines are blended in the G395M grating.
Therefore, we corrected the [\ion{Ar}{4}] \W4713 flux for the 
\ion{He}{1} \W4714 contribution, predicting the \ion{He}{1} \W4714 
flux from the measured \ion{He}{1} \W4473 flux and the theoretical 
\ion{He}{1} \W4714/\W4473 ratio ($\sim0.21$ for the conditions in RXCJ2248-ID).
The resulting densities, all of which fall within their respective 
diagnostic ranges, and temperatures are reported in Table~\ref{tbl2}.

Remarkably, RXCJ2248-ID3 is one of few galaxies, and the only 
galaxy yet at high-redshifts, to have significant ($>3\sigma$) 
electron density measurements from five different ions that span 
a large ionization range ($\sim10-77$ eV).
Furthermore, the densities in RXCJ2248-ID3 appear to be organized into an 
interesting nebular stratification.
The UV emission lines trace the densest gas, with 
$n_e$(N$^{+3})=\nNIV\times10^5$ cm$^{-3}$ in the highest ionization gas, 
followed by $n_e$(C$^{+2})=\nCIII\times10^4$ cm$^{-3}$, and 
$n_e$(Si$^{+2})=\nSiIII\times10^4$ cm$^{-3}$.
In contrast, the optical high ionization lines are emitted from regions 
of lower densities: the optical [\ion{Ar}{4}] diagnostic has an overlapping 
ionization energy range with the UV \ion{N}{4}] diagnostic, but a density 
that is an order of magnitude lower.

There are two possible interpretations of the measured array of densities.
First, since the UV lines also have higher excitation energies, they could 
originate preferentially from hotter, denser clumps. 
This would imply a strongly inhomogeneous ISM, in which compact, high-pressure 
structures dominate the UV line emission while somewhat more diffuse 
gas produces much of the optical emission. 
Alternatively, the multiphase ISM may span a smaller dynamic range of 
densities than we measure due to the suppression of the optical diagnostics.
\citet{martinez25} showed that for an ISM with a mix of low- (e.g., $10^3$ cm$^{-3}$)
and high-density gas (e.g., $10^5$ cm$^{-3}$) that has a true volumetric
density that is somewhere in between, the low-ionization optical diagnostics will always 
be {\it significantly} biased low, close to the low-density gas value, until the fraction 
of high-density gas is very high (e.g., $>95\%$).
This effect occurs when $n_e$-diagnostic line ratios have low-critical densities (e.g., 
$n_{e,crit}$([\ion{S}{2}])$\approx2\times10^3-5\times10^3$ cm$^{-3}$), such that emission
from the high-density gas is collisionally-suppressed beyond detection.
The magnitude of this effect decreases with increasing critical density such
that [\ion{S}{2}] is significantly affected, [\ion{Ar}{4}] is moderately affected 
($n_{e,crit}\approx2\times10^4-2\times10^5$ 
cm$^{-3}$), and the UV \ion{Si}{3}], \ion{C}{3}], and \ion{N}{4}] 
($n_{e,crit}\approx5\times10^4-5\times10^{10}$ cm$^{-3}$)
are minimally affected, albeit still biased low.
In this scenario, there would still be density stratification, but with 
smaller differences. \looseness=-2

All together, the nebular diagnostics in RXCJ2248-ID3 support a picture of a 
multiphase nebula with density and temperature stratification, 
likely reflecting a clumpy ISM shaped by the feedback and local radiation field 
variations of bursty star formation \citep[see, also,][]{harikane25,usui25,choustikov25}.
This picture is also consistent with the density stratification that has 
been reported for dwarf galaxies both near and far 
\citep[e.g.,][]{james16,berg21,mingozzi22,ji24,topping24a}, but with
typical densities increasing with redshift 
\citep[e.g.,][]{isobe23,abdurrouf24,topping25b,martinez25}. 

\subsection{Ionization Parameter}\label{sec:logu}
The ionization parameter of RXCJ2248-ID3, $\log U$, determined using the typical 
O$_{32} = I_{\lambda5008}/I_{\lambda3728}$ diagnostic is reported in 
\citet{topping24a} to be in the high range of $\log U = -0.5$ to $-1$.
We recompute the ionization parameter for RXCJ2248-ID3 using the O$_{32}$ and 
N$_{43} = I_{\lambda\lambda1483,1486}/I_{\lambda1750}$ diagnostics 
from \citet{martinez25} that are calibrated for densities in the 
$10^2 \leq n_e\ ({\rm cm}^{-3}) \leq 10^6$ range.
We estimate a $\log U_{\rm int} = \lui\pm0.23$ using O$_{32}$, which is 
consistent with the value reported by \citet{topping24a}, and 
$\log U_{\rm high} = \luh\pm0.23$ using N$_{43}$. 
Note, however, that the O$_{32}$ diagnostic is very sensitive to the 
assumed density \citep{martinez25}, making this value highly uncertain in
dense gas.
For example, densities of $n_e = 10^3$ to $10^5$ cm$^{-3}$ would lead to a range of 
$\log U_{\rm int} = -0.57$ to $-2.04$, respectively.\looseness=-2 

\subsection{Abundances}\label{sec:abund}
Here we present direct-method abundances of O/H (\S~\ref{sec:OH}), 
N/O (\S~\ref{sec:NO}), C/O (\S~\ref{sec:CO}), and Si/O (\S~\ref{sec:SiO}) 
for RXCJ2248-ID3 using narrow-line flux ratios and the measured 
temperatures and densities presented in \S~\ref{sec:temden}.
Nearly all of the optical lines used in this work have sufficient S/N to 
to simultaneously constrain broad and narrow emission components, but there are a 
few exceptions, all of which are low-ionization lines.
The [\ion{O}{2}] \W3728 line was not covered by the GLIMPSE-D spectrum and 
so lacks the S/N to fit broad components.
Both [\ion{N}{2}] \W6585 and [\ion{S}{2}] \W\W6718,6733 are covered in the high-S/N
GLIMPSE-D spectrum but are either blended with stronger features or too weak to fit 
broad components.
On the other hand, the [\ion{O}{2}] and [\ion{S}{2}] lines have low critical 
densities around $n_{e,crit}\sim 10^3{\rm\ cm}^{-3}$ such that any moderate to 
high density broad components are likely collisionally de-excited away. 
Their narrow component fluxes could also be significantly reduced by collisional 
de-excitation, however, the missing [\ion{O}{2}] emission is likely small in the 
absolute sense for such a high-ionization object. 
Emission from [\ion{N}{2}] is less likely to be collisionally de-excited 
($n_{e,crit}\sim 10^5{\rm\ cm}^{-3}$), so a hidden broad component could lead
to an overestimate of the N/O abundance, but this effect would be somewhat 
countered by the underestimated [\ion{O}{2}] flux.
In the end, the consistency of N/O derived independently from UV and optical tracers 
in \S~\ref{sec:NO} below suggests that these effects do not significantly
impact our results.

To calculate the total or relative abundance of an element, 
we determine and sum the individual observed ions and then apply an 
ionization correction factor to account for unseen prominent ionization states. 
The abundance of an individual ionic species, $X^i$, relative to hydrogen is 
determined as:
\begin{equation}
    \frac{N(X^i)}{N(H^+)} = \frac{I_{\lambda(i)}}{I_{\rm H\beta}} 
                            \frac{j_{\rm H\beta}}{j_{\lambda(i)}},
\end{equation}
where $j_{\lambda(i)}$ is the emissivity determined for the
appropriate ionization zone temperature and density. 
Given the tendency of the optical density diagnostics to severely under-estimate
the density in high-density environments, we instead adopt the UV-derived densities.
Note that abundances presented below have not been corrected for the fraction 
of atoms embedded in dust. 
However, the level of depletion onto dust grains is expected to be small for 
the low metallicity of RXCJ2248-ID3 
\citep[e.g.,][]{remy-ruyer14,galliano18,roman-duval22}.
\citet{isobe26} also infer negligible dust depletion for RXCJ2248-ID3 
based on the high value of Si/O that that they determine, 
but this is inconsistent with the value we determine below.
Details of elemental abundance determinations are given below.

\subsubsection{Oxygen Abundance}\label{sec:OH}
We determine the total O/H abundance as the sum of the O$^+$/H$^+$ and
O$^{+2}$/H$^+$ ionic abundances, determined from the [\ion{O}{2}] \W3728
and [\ion{O}{3}] \W\W4960,5008 optical emission lines. 
We observe no strong O$^0$ or O$^{+3}$ emission, indicating
contributions from other ions are negligible. 
The resulting ionic and total oxygen abundances are presented in Table~\ref{tbl2}.
Similar to \citet{hayes25} and \citet{martinez25}, we find that one of the most 
significant effects of accounting for high densities is the resulting decrease in 
electron temperature and subsequent increase in oxygen abundance
\citep[see, also,][]{katz23}.
In our work, this results both from accounting for the missing [\ion{O}{3}]
\W5008 flux due to collisional de-excitation and from correcting the narrow
emission for broad emission components at their base.
\citet{topping24a} also incorporated the high densities seen in RXCJ2248-ID3,
but did not have the S/N to fit the broad emission components in both
[\ion{O}{3}] \W5008 and \W4364. 
As a result, we measure an oxygen abundance of $12+\log(\rm O/H)=\loh\pm\lohe$.
Note that if unresolved high-density clumps ($n_e\gtrsim10^5\ {\rm cm}^{-3}$) 
are present (as suggested in, e.g., \S~\ref{sec:red}), it could introduce additional 
uncertainty by biasing the luminosity-weighted [\ion{O}{3}] \W4364/\W5008 ratio to 
higher densities, which would drive the derived $T_e$ higher and O/H abundance lower. 
However, \citet[][see Figure 11]{martinez25}, showed that the use of high-critical 
density UV density diagnostics largely mitigate this effect in a density stratified 
medium.

\subsubsection{Relative N/O Abundance}\label{sec:NO}
The extraordinary simultaneous detections of [\ion{N}{2}] \W6585, \ion{N}{3}] \W1750,
and \ion{N}{4}] \W\W1483,1487 enable multiple determinations of the N/O abundance.
Therefore, we calculate N/O abundances using four different ionic methods:
\begin{align}
        1.\ \frac{\rm N}{\rm O} &= \frac{\rm N^+}{\rm O^+} \times {\rm ICF}=\frac{\rm N^+}{\rm O^+}\times \left[\frac{X(\rm N^+)}{X(\rm O^+)}\right]^{-1} \\
        2.\ \frac{\rm N}{\rm O} &= \frac{\rm N^{+2}}{\rm O^{+2}} \times {\rm ICF}=\frac{\rm N^{+2}}{\rm O^{+2}} \left[\frac{X(\rm N^{+2})}{X(\rm O^{+2})}\right]^{-1} \\
        3.\ \frac{\rm N}{\rm O} &= \frac{\rm N^{+3}}{\rm O^{+2}} \times {\rm ICF}=\frac{\rm N^{+3}}{\rm O^{+2}} \left[\frac{X(\rm N^{+3})}{X(\rm O^{+2})}\right]^{-1} \\
        4.\ \frac{\rm N}{\rm O} &= \frac{\rm N^{+} + N^{+2} + N^{+3}}{\rm O^{+} + O^{+2}},
\end{align}
where [\ion{O}{2}] \W3728 is used for the N$^{+}$/O$^{+}$ determination, 
\ion{O}{3}] \W1666 is used for the N$^{+2}$/O$^{+2}$ and N$^{+3}$/O$^{+2}$ calculations, 
and $X(\rm N^{+i})$ and $X(\rm O^{+i})$ are the N and O ionization fractions, respectively.
We use the density-dependent ICFs 
from \citet{martinez25}, who provide prescriptions for densities of 
$n_e = 10^2, 10^3, 10^4, 10^5,$ and $10^6$ cm$^{-3}$.
We, therefore, round our density measurements to the nearest order of magnitude
and use the intermediate-ionization $\log U$ for the N$^{+}/$O$^{+}$ ICF
and the high-ionization $\log U$ for the N$^{+2}/$O$^{+2}$ and N$^{+3}/$O$^{+2}$ ICFs.
The resulting N ICFs and N/O abundances are reported in Table~\ref{tbl3}.

The four N/O determinations of RXCJ2248-ID3 are in close agreement,
far above the expected value for its metallicity. 
Visually, this is shown in the upper left-hand panel of Figure~\ref{fig:CNO},
which plots the relative N/O versus O/H abundance with RXCJ2248-ID3 
marked by purple diamonds.
The traditional N/O--O/H trend has been established by many $z\sim0$ 
studies of \ion{H}{2} regions and galaxies 
\citep[gray points:][]{esteban02,pilyugin05,vanzee06a,garcia-rojas07,lopez-sanchez07,esteban09,berg12,esteban14,berg16,berg19a,berg20}.
The empirical trend is a bimodal relationship, with a flat trend due to 
primary (or metallicity-independent) N production at low-metallicities 
($12+\log(\rm O/H)\lesssim8.0$) and an increasing N/O trend with O/H 
as secondary (or metallicity-dependent) N production becomes 
increasingly important at higher metallicities ($12+\log(\rm O/H)\gtrsim8.0$).
As a visual guide, the primary N/O plateau from \citet[][dashed purple line]{berg19a} is 
shown and the empirical stellar curve from \citet[][solid green line]{nicholls17} 
is shown as an example of the full primary+secondary curve.

For comparison, we plot the high-quality high-redshift N/O measurements that were 
calculated in a consistent manner as the present work (with direct-method $T_e$ 
and $n_e$ determinations and $n_e$-dependent ICFs) by \citet{martinez25}.
N/O abundances determined using N$^{+2}$/O$^{+2}$ are plotted as blue $+$ symbols, 
while N$^{+}$/O$^{+}$ determinations are plotted as blue pentagons. 
Of these galaxies, the closest comparison to RXCJ2248-ID3 is CEERS-1019 
\citep[see, also,][]{marques-chaves24}, while only GDS 3073 and GN-z11 have 
higher relative N/O abundances and only GDS 3073 is more enhanced in N/O for 
its O/H abundance.

We find that all four ionic methods produce consistently high 
N/O values within their uncertainties, with a weighted mean of 
$\log({\rm N/O})=\lnoave\pm\lnoavee$.
This is an important result, as RXCJ2248-ID3 is the first galaxy to have 
consistently enhanced N/O abundances measured from both the rest-frame UV 
high-ionization and the optical low-ionization emission lines. 
Furthermore, measuring consistent N/O values from three different ionic methods
strengthens our confidence in the robustness of the N/O measurement,
although uniform N/O across the ionization structure of the nebula is 
not a given in a stratified medium.
While there is strong evidence for a stratified, or perhaps very clumpy, 
density structure in RXCJ2248-ID, the N/O abundance appears to be well mixed. 

\subsubsection{Relative C/O Abundance}\label{sec:CO}
Measuring the carbon-to-oxygen (C/O) abundance provides a crucial comparative 
baseline for interpreting the origin of elevated N/O in RXCJ2248-ID3. 
Similar to N, C has a pseudo-secondary\footnote{There are no know secondary
nucleosynthetic production methods for C; only the primary triple-$\alpha$
process is known to produce C. However, other processes, such as metal-dependent 
line-drive stellar winds, may be responsible for the empirical increasing trend 
in C/O--OH, hence the term {\it pseudo-secondary}.} production pathway, 
but the dominant nucleosynthetic sources and timescales differ for C and N. 
Briefly, both C and O are primarily produced in massive stars ($>8\ M_\odot$) 
on relatively short timescales such that the C/O ratio is a relatively stable 
tracer of massive star yields, although some C is produced via low- to 
intermediate-mass AGB stars ($\sim1.5-3\ M_\odot$).
In contrast, some N is produced by massive stars (e.g., through rotational mixing 
and WR winds) but most N comes from intermediate-mass AGB stars 
($\sim4-8\ M_\odot$), which release N on longer timescales ($\sim200$ Myr). 
Therefore, N/O and C/O together serves as diagnostics of the recent star formation
history, constraining the recent enrichment mechanisms of galaxies 
\citep[e.g.,][]{garnett90,henry00,chiappini03,perez-montero09,berg19a,perez-montero21}.

Relative C/O abundances are typically determined using the 
\ion{C}{3}] \W\W 1907,1909/\ion{O}{3}] \W1666 ratio to calculate C$^{+2}$/O$^{+2}$ 
and assuming that C/O $\approx$ C$^{+2}$/O$^{+2}$.
This method is sometimes used alone owing to the fact that 
(1) C$^{+2}$ and O$^{+2}$ have somewhat similar ionization potentials 
(24.38 and 35.12 eV, respectively),
(2) the upper levels of the \W1666 and \W\W1907,1909 transitions have similar 
excitation potentials ($\sim$6.5 and $\sim$7.5 eV, respectively), and 
(3) the integrated fluxes of \W1666 and \W\W1907,1909 are not sensitive to 
collisional de-excitation for the densities measured here.
However, for the high-ionization nebulae in RXCJ2248-ID3, it is important to 
account for contributions from the C$^{+3}$ species and any unseen species. 
We note that the \ion{C}{4} \W\W1548,1550 doublet is clearly observed in the 
rest-UV spectrum of RXCJ2248-ID3, but these lines are resonant and can be affected 
by the \ion{C}{4} stellar wind feature and ISM absorption, and so, determining 
the intrinsic flux and subsequent C$^{+3}$ abundance is challenging.
Instead, we use an ICF determined from the photoionization models presented in 
\citet{martinez25} such that 
\begin{equation}
    \frac{\rm C}{\rm O} =
    \frac{\rm C^{+2}}{\rm O^{+2}}\times{\rm ICF} =
    \frac{\rm C^{+2}}{\rm O^{+2}}\times\left[\frac{X(\rm C^{+2})}{X(\rm O^{+2})}\right]^{-1}.
\end{equation}
We used the $\log U_{\rm int.}$ and a density of 
$n_e(\rm C^{+2})\sim10^5$ cm$^{-3}$ to determine the C ICF.
The resulting C ICF and C/O abundance are reported in Table~\ref{tbl3}.

The C/O and C/N abundances for RXCJ2248-ID3 are plotted in the upper middle 
and right-hand panels of Figure~\ref{fig:CNO}.
Empirical trends of C/N at $z\sim0$ are found to be flat, albeit with 
significant scatter (see shading in Figure~\ref{fig:CNO}), suggesting that 
the dominant nucleosynthetic mechanisms of C are similar to those of 
N \citep[e.g.,][]{garnett99,esteban14,berg16,berg19a}.
However, while the production of both C and N appear to be metallicity-dependent, 
the scatter in their trend is consistent with differing production timescales 
due to stars of different masses. 
Thus, the variations observed in CNO abundance patterns of high-redshift galaxies 
may be the result of taking a snapshot of many galaxies at different times since 
their most recent onset of star formation.

RXCJ2248-ID3 appears to have a similar CNO abundance pattern to other high-redshift 
N-emitters, characterized by enhanced N/O but relatively deficient C/O such that 
their C/N is very deficient compared to the expectations from low-redshift trends.
This suggests that these high-redshift N-emitting galaxies are enhanced in 
N relative to both O and C.
If massive stars in the WN phase are present, they will have recently produced 
$^{14}$N at the expense of $^{12}$C through the CNO cycle, meaning 
C used as a catalyst in the cycle initiation will have been {\it consumed} as 
N is removed during the bottleneck step via dredged up, preventing the return of 
C at cycle completion. 
Thus, C/N-deficiency is consistent with a recent, intense episode of N-enrichment 
and C-consumption from WN stars.
Conversely, if both N/O and C/O were elevated in tandem, it could point to broader 
enrichment by massive stars, such as enrichment from both WN {\it and} WC stars, 
whose contributions increase at higher metallicities.

\subsubsection{Relative Si/O Abundance}\label{sec:SiO}
Detecting \ion{Si}{3}] \W\W1883,1892 in RXCJ2248-ID3 enables the rare 
opportunity to measure the silicon-to-oxygen (Si/O) abundance in a $z>5$ galaxy
\citep[see, also,][for Si/O in GN-z11]{isobe26}.
Silicon abundances are important for multiple reasons.
Silicon is highly refractory, making the Si/O ratio a sensitive probe of dust depletion.
Additionally, Si probes different channels of chemical enrichment than CNO elements, 
as it is primarily an $\alpha$-element produced by CCSNe, but Type Ia SNe, AGB stars, 
and even pair-instability SNe are all expected to contribute to the total
Si abundance.
For RXCJ2248-ID3 we determine the Si/O abundance using the observed \ion{Si}{3}] 
\W\W 1883,1892/\ion{O}{3}] \W1666 ratio to calculate Si$^{+2}$/O$^{+2}$.
Because Si$^{+2}$ and O$^{+2}$ have rather different ionization potentials
(16.3 eV versus 35.1 eV, respectively), a Si ICF is required to convert 
Si$^{+2}$/O$^{+2}$ to total Si/O via
\begin{equation}
    \frac{\rm Si}{\rm O} =
    \frac{\rm Si^{+2}}{\rm O^{+2}}\times{\rm ICF} =
    \frac{\rm Si^{+2}}{\rm O^{+2}}\times\left[\frac{X(\rm Si^{+2})}{X(\rm O^{+2})}\right]^{-1}.
\end{equation}
Si ICFs have been reported previously \citep[e.g.,][]{garnett95b}, but 
none account for the high density conditions observed in RXCJ2248-ID3.
Therefore, we determined a Si ICF $=3.507$ using from the photoionization models 
presented in \citet{martinez25} using the $\log U_{\rm int.}$ and a density of 
$n_e(\rm Si^{+2})\sim10^4$ cm$^{-3}$.
Reported in Table~\ref{tbl3}, the resulting $\log$(Si/O) $=-1.781\pm0.157$ 
abundance is typical of metal-poor dwarf galaxies \citep[e.g.,][]{garnett95b,izotov99}, 
consistent with normal massive star production and low dust depletion.

\subsection{Potential Impact of UV Broad Components}\label{sec:broad_abund}
The exceptionally high S/N of the rest-optical GLIMPSE-D spectrum allows for 
broad emission component fits that the rest-UV spectrum does not. 
In Section~\ref{sec:em} we found the broad emission component contribution
to the narrow [\ion{O}{3}] \W5008 flux to be 10.8\%. 
To examine the possible effects such contamination has on calculations of
nebular conditions and abundances, we adopt 10.8\%\ as the contamination 
upper limit to the UV emission lines. 
We first consider the impact on the UV density determinations, where we 
allow the broad components of the UV density-sensitive emission line ratios
to have densities ranging from $10^2-10^6\ {\rm cm}^{-3}$. 
After subtracting the potential broad component contribution, the revised
densities change up to 
$\Delta n_e$(Si$^{+2}$)$=^{+0.34}_{-0.67}\times10^4 {\rm\ cm}^{-3}$,
$\Delta n_e$(C$^{+2}$)$=^{+0.78}_{-0.85}\times10^4 {\rm\ cm}^{-3}$, and
$\Delta n_e$(N$^{+3}$)$=^{+0.32}_{-0.12}\times10^5 {\rm\ cm}^{-3}$ 
over the range of broad component densities considered.
These values are within the reported uncertainties in Table~\ref{tbl3},
with the exception of the $\sim1.2\sigma$ deviation for $\Delta n_e$ of C$^{+2}$.

Next, we tested the subsequent impact of UV densities that have been revised for 
possible broad components on the properties determined from rest-optical 
emission lines: $T_e({\rm O}^{+2})$, O/H, and N/O.
For the range of $\Delta n_e$(N$^{+3}$) above, the resulting 
$\Delta T_{e,{\rm high}} = ^{+0.029}_{-0.064}\times10^4$ K, which is within
$1-2\sigma$ of the reported value in Table~\ref{tbl3}.
Similarly, the impact of the revised densities and temperatures on the 
oxygen abundance, $\Delta{\rm O/H}=^{+0.046}_{-0.017}$ dex, is also within
$1-2\sigma$.
The impact is even smaller for the nitrogen abundance, with 
$\Delta{\rm (N/O)_{N^+}}=^{+0.024}_{-0.038}$ dex being much smaller than the 
N/O uncertainty. 

Relative UV abundances are impacted by changes in both the nebular conditions 
and the relevant abundance emission line ratio.
However, the resulting abundance deviations are small and within the 
original uncertainties:
$\Delta{\rm (N/O)_{N^{+2}}}=^{+0.005}_{-0.002}$ dex,
$\Delta{\rm (N/O)_{N^{+3}}}=^{+0.003}_{-0.001}$ dex, and
$\Delta{\rm (C/O)}=^{+0.011}_{-0.003}$ dex.
Thus, we conclude that while considering the impacts of hidden broad 
component contributions to the measured UV fluxes is important, the
potential biases do not affect the main results and conclusions of this work. 


\begin{deluxetable}{rCcc}
\setlength{\tabcolsep}{5pt}
\tablewidth{0pt}
\tablecaption{Nebular Conditions and Abundances for RXCJ2248-ID3}
\tablehead{
\CH{Property}    & \CH{Ion. E (eV)} & \CH{Used} & \CH{Value}}\startdata \\[-1ex]
{\bf Temperatures:} & & \\
$T_{e,{\rm high}}$ meas. (K) & 35.11-54.93   & $n_e$(N$^{+3}$) & $1.97\pm0.03\times10^4$  \\
$T_{e,{\rm int.}}$ used (K)  & 23.33-34.83   & \citet{garnett92} & $1.81\pm0.02\times10^4$  \\
$T_{e,{\rm low}}$ used (K)   & 13.62-35.11   & \citet{garnett92} & $1.68\pm0.02\times10^4$              
\vspace{2ex} \\
{\bf Densities:} &              &   &                                            \\
$n_e$(N$^{+3}$) (cm$^{-3}$)  & 47.45-77.47  & $T_{e,{\rm high}}$ & $2.65^{+0.45}_{-0.38}\times10^5$  \\  
$n_e$(Ar$^{+3}$) (cm$^{-3}$) & 40.74-59.81  & $T_{e,{\rm high}}$ & $1.75^{+0.26}_{-0.22}\times10^4$ \\
$n_e$(C$^{+2}$) (cm$^{-3}$)  & 24.38-47.89  & $T_{e,{\rm int.}}$ & $7.94^{+0.69}_{-0.65}\times10^4$ \\
$n_e$(Si$^{+2}$) (cm$^{-3}$) & 16.35-33.49  & $T_{e,{\rm int.}}$ & $4.77^{+5.68}_{-2.33}\times10^4$ \\
$n_e$(S$^{+}$) (cm$^{-3}$) 	 & 10.36-23.33  & $T_{e,{\rm low}}$  & $1.15^{+2.23}_{-0.69}\times10^3$ 
\vspace{2ex} \\
\multicolumn{4}{l}{\bf O Abundances:} \\
O$^+$/H$^+$ ($\times 10^{-5}$)   & 13.62-35.11 & $T_{e,{\rm low}};\ n_e$(Si$^{+2}$)   & $0.186\pm0.148$ \\
O$^{+2}$/H$^+$ ($\times 10^{-5}$)& 35.11-54.93 & $T_{e,{\rm high}};\ n_e$(Ar$^{+2}$)  & $5.473\pm0.261$ \\
$12+\log(\rm O/H)$ 	 	 	 	 &             &                                      & $7.753\pm0.023$  
\vspace{2ex} \\
\multicolumn{4}{l}{\bf Ionization Parameters:} \\
log$U_{\rm int.}$(O$_{32}$)     & 13.62-54.93  & $n_e=10^4$ cm$^{-3}$                 & $-1.24\pm0.23$ \\
log$U_{\rm high}$(N$_{43}$)     & 29.60-77.47  & $n_e=10^5$ cm$^{-3}$                 & $-0.69\pm0.10$       
\vspace{2ex} \\
\multicolumn{4}{l}{\bf N Abundances:} \\
N$^{+3}$/O$^{+2}$ 	 	 	 	& 47.45-77.47   & $T_{e,{\rm high}};\ n_e$(N$^{+3}$)  & $0.277\pm0.043$  \\
N$^{+2}$/O$^{+2}$ 	 	 	 	& 29.60-47.45   & $T_{e,{\rm high}};\ n_e$(C$^{+2}$)  & $0.145\pm0.070$  \\
N$^+$/O$^+$ 	 	 	 	 	& 14.53-29.60   & $T_{e,{\rm low}};\ n_e$(Si$^{+2}$)  & $0.367\pm0.259$  \\
ICF(N$^{+3}$/O$^{+2}$) 	   	 	& 47.45-77.47   & $T_{e,{\rm high}};\ n_e$(N$^{+3}$)  & 1.542 \\
ICF(N$^{+2}$/O$^{+2}$)          & 29.60-47.45   & $T_{e,{\rm high}};\ n_e$(C$^{+2}$)  & 2.547 \\
ICF(N$^{+}$/O$^{+}$)            & 14.53-29.60   & $T_{e,{\rm low}};\ n_e$(Si$^{+2}$)  & 0.814 \\
log(N/O)$_{\rm N^{+3}}$         &               &   & $-0.368\pm0.062$ \\
log(N/O)$_{\rm N^{+2}}$         &               &   & $-0.434\pm0.071$ \\
log(N/O)$_{\rm N^+}$            &               &   & $-0.525\pm0.257$ \\
log(N/O)$_{\rm all}$            &               &   & $-0.375\pm0.056$ \\
$\left<\log(\rm N/O) \right>$   &               &   & $-0.390\pm0.035$ 
\vspace{2ex} \\
\multicolumn{4}{l}{\bf C Abundance:} \\
C$^{+2}$/O$^{+2}$ 	 	 	 	& 24.38-47.89   & $T_{e,{\rm int}};\ n_e$(C$^{+2}$)   & $0.107\pm0.014$  \\
ICF(C$^{+2}$/O$^{+2}$)          & 24.38-47.89   & $T_{e,{\rm int}};\ n_e$(C$^{+2}$)   & 1.498 \\
log(C/O)                        &               &                                     & $-0.795\pm0.052$ 
\vspace{2ex} \\
\multicolumn{4}{l}{\bf Si Abundance:} \\
Si$^{+2}$/O$^{+2}$ 	 	 	 	& 16.35-33.49   & $T_{e,{\rm low}};\ n_e$(Si$^{+2}$)  & $0.005\pm0.001$  \\
ICF(Si$^{+2}$/O$^{+2}$)         & 16.35-33.49   & $T_{e,{\rm low}};\ n_e$(Si$^{+2}$)  & 3.507 \\
log(Si/O)                       &               &                                     & $-1.781\pm0.157$ 
\enddata
\tablecomments{Ionic and total abundances for RXCJ2248-ID3.
Column 1 lists the property, while Column 2 lists the associated ionization potential
energy range (eV), Column 3 lists the temperature and/or density used in the calculation,
and Column 4 provides the final values.
All calculations reported here only used the narrow components when multicomponent
fits were performed.
Note that the temperatures for the intermediate- and low-ionization zones 
were inferred from $T_{e,{\rm high}}$ using the $T_e-T_e$ relationships 
of \citet{garnett92}.
Two ionization parameters are reported for the O$_{32}$ and N$_{43}$ indicators
from \citet{martinez25}.
The oxygen abundance was determined using the archival [\ion{O}{2}] \W3728 detection and 
the new [\ion{O}{3}] \W5008 fit.
N/O was determined using four different ion$+$ICF \citep[from][]{martinez25} combinations: 
(1) optical N$^{+}$/O$^{+}$; 
(2) UV N$^{+2}$/O$^{+2}$; 
(3) UV N$^{+3}$/O$^{+2}$; and
(4) combination (N$^{+}$+N$^{+2}$+N$^{+3}$)/(O$^{+}$+O$^{+2}$).
C/O and Si/O were determined from the archival UV emission lines only. 
}\label{tbl3}
\end{deluxetable}


\begin{figure*}
\begin{center}
	\includegraphics[width=1.0\linewidth,trim=0mm 6mm 0mm 10mm,clip]{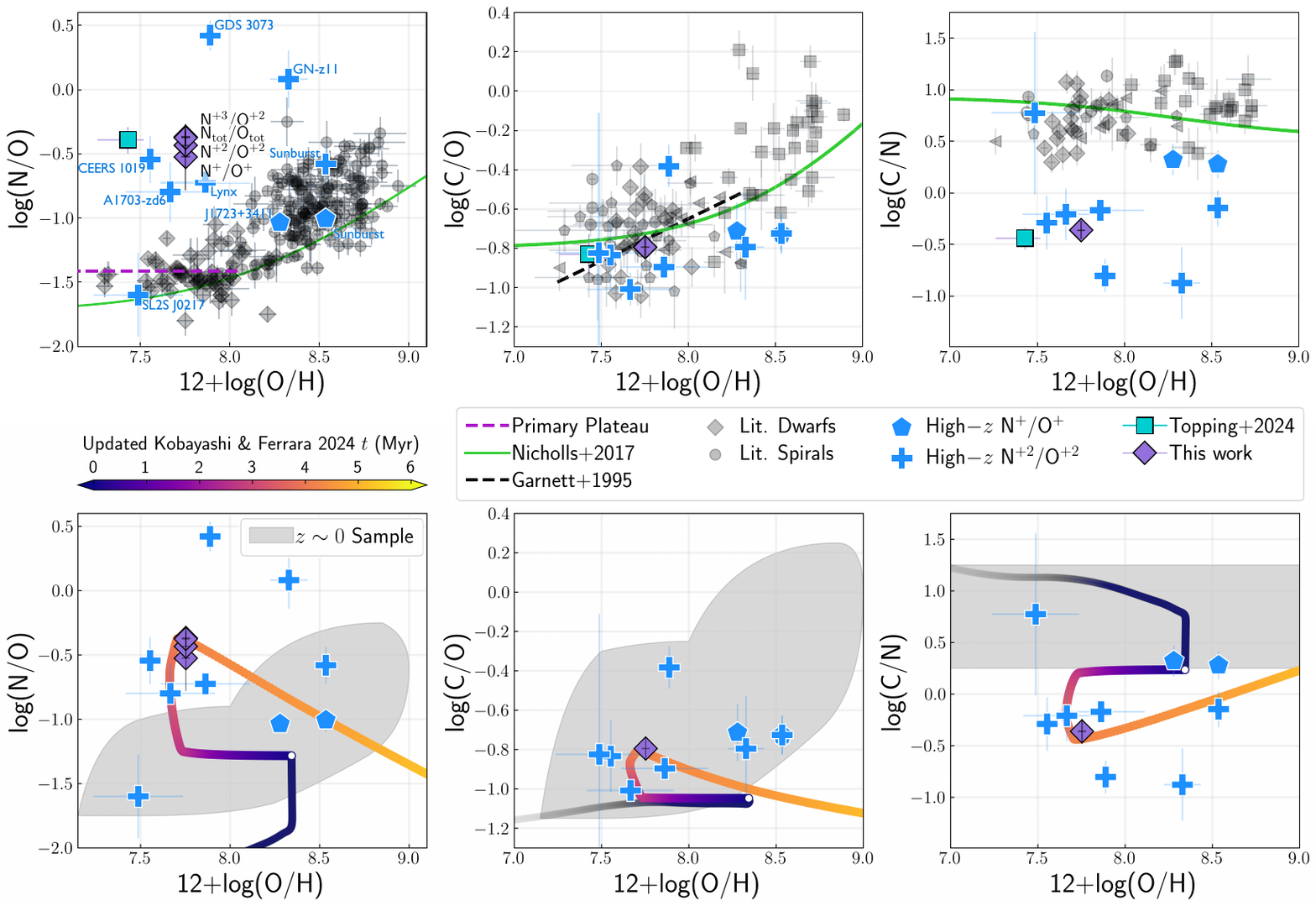} 
\caption{Relative C and N abundance trends versus metallicity.
Nitrogen to oxygen ratio versus oxygen abundance for star-forming galaxies is
plotted in the left panels, while carbon to oxygen ratio versus oxygen abundance is
plotted in the middle panels, and carbon to nitrogen abundance versus oxygen 
abundance is plotted in the right panels.
{\it Top row:} 
RXCJ2248-ID3 is shown relative to the observed $z\sim0$ trend and other high-$z$ 
galaxies.
The abundances for RXCJ2248-ID3 are shown as purple diamonds, where multiple N/O 
points show the measurements for each ionic N/O calculation method.
For comparison, we also plot the abundances derived for RXCJ2248-ID3 by \citet{topping24a} as turquoise squares.
The typical bi-modal N/O trend is characterized by 
local dwarf \citep[gray diamonds;][]{vanzee06a,berg12,berg16,berg19a} and 
spiral galaxy \citep[gray circles;][]{esteban02,esteban09,esteban14,pilyugin05,garcia-rojas07,lopez-sanchez07,berg20} \ion{H}{2} region measurements.
The primary N/O plateau from \citet{berg19a} is shown as a dashed purple line,
while the solid green line is the empirical stellar curve from \citet{nicholls17}.
Additional C/O literature measurements for dwarf galaxies are from 
\citet{pena-guerrero17} and \citet{senchyna17}.
Abundances for $z>2$ galaxies from \citet{martinez25} are plotted as blue plus 
signs for galaxies with UV N$^{+2}$/O$^{+2}$ derived abundances and pentagons for 
optical N$^+$/O$^+$ derived abundances.
{\it Bottom row:} 
The same observed samples are shown as the top row, but with the $z\sim0$ sample
represented by the shaded gray regions.
The observed abundances of RXCJ2248-ID3 are compared to updated dual-burst 
chemical evolution models of \citet[][string of circles]{kobayashi24}, 
color-coded by age since onset of the second burst.
The models have been modified to reproduce both the enhanced N/O and 
relatively deficient C/O observed for RXCJ2248-ID3, which requires 
enrichment from WN but very little WC enrichment, as expected at low-metallicities.
{ \label{fig:CNO}}}
\end{center}
\end{figure*}

\section{A Short Window of Intense WR Nitrogen Enrichment}\label{sec:source}
We have presented evidence for WN stars in RXCJ2248-ID3 in two forms:
first, the rest-frame optical WR blue bump discussed in Section~\ref{sec:WR} and 
shown in Figure~\ref{fig:spec}; second, a qualitative comparison of the CNO 
abundances to patterns expected for WR stars in \S~\ref{sec:abund} and Figure~\ref{fig:CNO}.
Below, we examine the plausibility and impact of these WN stars by 
comparing RXCJ2248-ID3 to expected trends for WR stars with metallicity (\S~\ref{sec:LZWN}),
testing whether stellar yields can reproduce the observed CNO abundance pattern 
(\S~\ref{sec:chemevol}), and assessing whether RXCJ2248-ID3's stellar population can 
produce its inferred mass of ionized N (\S~\ref{sec:Nmass}).
Together, these lines of investigation suggest that the enhanced N/O and 
suppressed C/O in RXCJ2248-ID3 represent a short-lived enrichment phase,
unique to metal-poor, highly star-forming galaxies in the early Universe 
(\S~\ref{sec:time}).

\subsection{WN Stars: The Dominant WR Phase at Low-Metallicity}\label{sec:LZWN}

\begin{figure}
\begin{center}
	\includegraphics[width=1.0\linewidth,trim=0mm 4mm 0mm 0mm,clip]{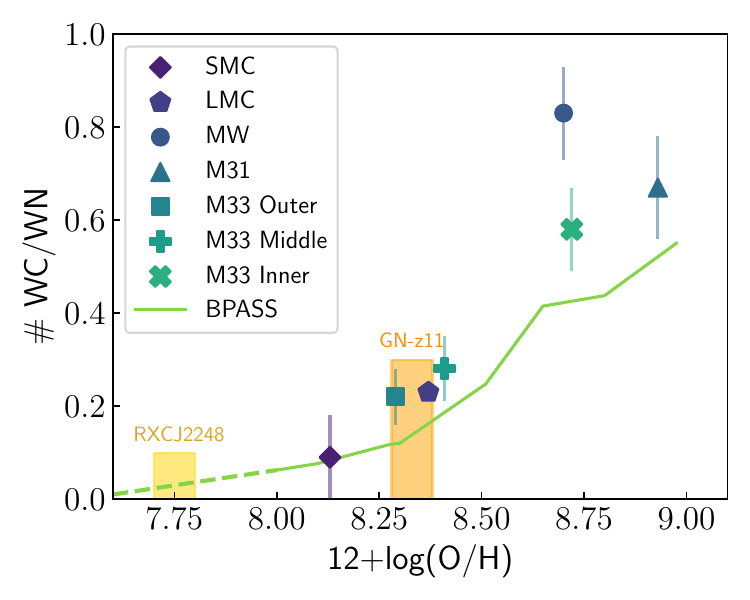} 
\caption{Observed and theoretical ratios of WC/WN star numbers as a function of metallicity.
Observed values for the SMC, LMC, Milky Way, and M31 were compiled by \citet{neugent19},
while newer values for M33 come from \citet{neugent23}. 
For comparison, we also plot the trend presented in \citet{massey17} for BPASSv2.0 binary 
stellar population synthesis burst models for 12+log(O/H) $>8$ (solid line green line),
which we extrapolate to lower metallicities (dashed line).
Enrichment from WR stars was use to explain the CNO abundances in GN-z11 
by \citet{kobayashi24}.
We note the metallicity for GN-z11 determined by \citet{martinez25}, which is consistent
with a WC/WN ratio of $\sim0.1-0.2$, and the metallicity for RXCJ2248-ID3 from the 
current work, which predicts a much lower WC/WN ratio of $\sim0.03-0.10$.
Therefore, very little carbon enrichment from WC stars is expected for 
RXCJ2248-ID3. \label{fig:WCN}}
\end{center}
\end{figure}


To date, no individual resolved WR stars have been directly observed at 
metallicities as low as RXCJ2248-ID3 ($Z\sim0.1Z_\odot$).
This is due, in part, to the lack of sufficiently close ($D\lesssim1$ Mpc for 
the young, crowded clusters hosting WR stars), metal-poor, star-forming galaxies 
(see \citealt{kehrig13} for the closest metal-poor WR galaxy), but 
a scarcity of WR stars in metal-poor environments is also expected 
because mass loss through stellar winds scales with metallicity.
We show the trend of the number of WC/WN star as a function of metallicity in 
Figure~\ref{fig:WCN}.
The observed number of WC/WN stars in M31 ($\sim175\%\ Z_\odot$), 
the Milky Way (MW; $Z_\odot$), 
M33 ($\sim40\%-110\%\ Z_\odot$),
the Large Magellanic Cloud (LMC; $\sim40\%\ Z_\odot$), and 
the Small Magellanic Cloud (SMC; $\sim20\%\ Z_\odot$) 
suggest that the number of WN/WC number ratio increases 
with decreasing metallicity \citep[e.g.,][]{meynet05,crowther07,massey15,neugent19}.
This is because weaker metal line-driven winds, rotation, or binary effects in 
metal-poor stars may be able to expose their nitrogen-rich layers and initiate 
the WN phase, but be insufficient to strip the stellar He atmosphere and reveal 
the carbon-rich core to initiate the WC phase.
Thus, if WR stars form at $Z \sim 10\%\,Z_\odot$, they are expected to be 
overwhelmingly WN-type.
Additionally, \citet{sander26} recently discovered a new class of
WN–WO stars that point to a low-metallicity WR evolutionary channel 
in which stars pass directly from the WN to WO phase, potentially explaining 
spectra that show evidence for WN-like enrichment and hard ionizing 
radiation without clear WC signatures.\looseness=-2

The spectral features of RXCJ2248-ID3 support the picture of 
WN star feedback at low metallicity.
As shown in Figure~\ref{fig:spec},
the \ion{He}{2} emission is moderately broadened, 
the \ion{N}{3} \W4642 line in the blue bump is prominent, and
there is no evidence for the red bump \ion{C}{4} feature, 
all consistent with the presence of WN stars at low metallicity.
The weakness of the \ion{He}{2} emission in terms of both flux and 
velocity width is expected for the low-metallicity environment 
of RXCJ2248-ID3 ($\sim10\%\ Z_\odot$) due to reduced wind velocities and 
mass-loss rates \citep[e.g.,][]{sander20}.
Similarly weak WN features have also been reported in the nearby metal-poor 
galaxy SBS 0335-052 \citep{izotov06b} and, 
at cosmic noon, the $z\sim2.37$ lensed galaxy the Sunburst Arc 
\citep{rivera-thorsen24} and the $z\sim2.22$ M4327 galaxy \citep{curti25b}.

\begin{figure}
\begin{center}
	\includegraphics[width=1.0\linewidth,trim=7mm 45mm 10mm 10mm,clip]{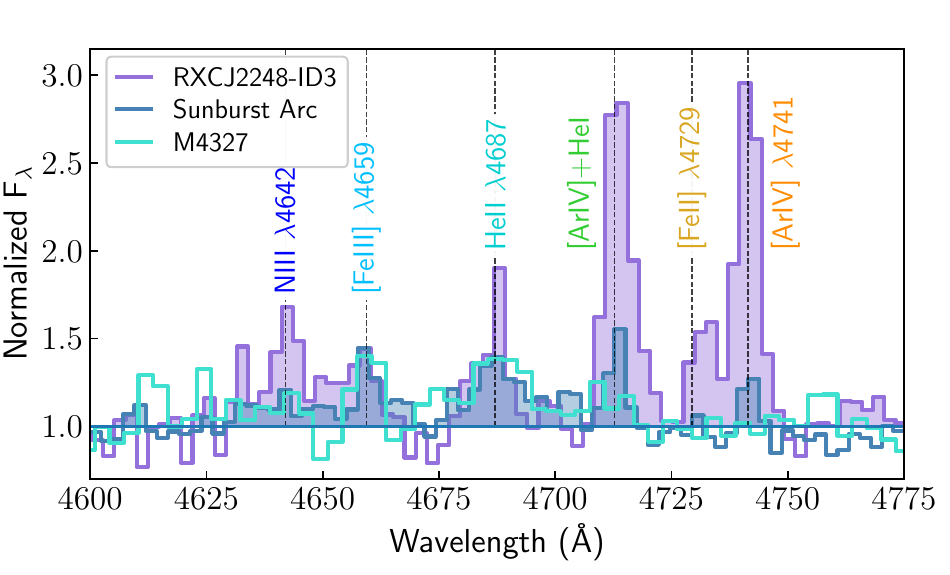}
\caption{
The blue WR region of the optical spectrum of RXCJ2248-ID3 (purple) is shown 
in comparison to the $z\sim2.37$ Sunburst Arc spectrum from 
\citet[][blue]{rivera-thorsen24}, 
which has been convolved to $R\sim1000$ to match RXCJ2248-ID3, and the $z=2.22$ M4327 
spectrum obtained from the DJA, but originally presented in \citet[][turquoise]{curti25b}.
All three galaxies show characteristic signs of hosting WN stars, but RXCJ2248-ID3
shows striking \ion{N}{3} \W4642 emission that is much stronger than both the Sunburst 
Arc and M4327.
On the other hand, the Sunburst Arc and M4327 show broader \ion{He}{2} emission, which 
is expected for more metal-rich galaxies as stellar winds scale with metallicity.
{\label{fig:WRspec}}}
\end{center}
\end{figure}

We plot the WR blue bump profile of RXCJ2248-ID3 relative to the Sunburst Arc 
and M4327 in Figure~\ref{fig:WRspec}.
For ease of comparison, we convolve the Sunburst Arc $R\sim2700$ JWST/NIRSpec 
G140H spectrum to the $R\sim1000$ resolution of the RXCJ2248-ID3 spectrum. 
For M4327, we retrieved the G140M spectrum obtained as part of the Measuring 
Abundance at high redshift with the $T_e$ Approach Survey \citep[MARTA;][]{cataldi25} 
from the Dawn JWST Archive \citep[DJA; ][]{heintz24,deGraaff25}. 
Both the Sunburst Arc and M4327 spectra were scaled to similar \ion{He}{2} 
strengths as RXCJ2248-ID3.
These spectra immediately reveal similar profiles, but with three distinct 
differences:
(1) RXCJ2248-ID3 exhibits higher gas ionization, as evidenced by the strong 
[\ion{Ar}{4}] \W\W4713,4741 emission;
(2) the \ion{He}{2} stellar wind feature is significantly broader in both the 
Sunburst Arc (FWHM$=1370$ km s$^{-1}$) and M4327 (FWHM$=1460$ km s$^{-1}$) 
than RXCJ2248-ID3 (FWHM$=\HeIIv$ km s$^{-1}$),
consistent with stronger stellar winds at the higher metallicities of the
Sunburst Arc: $12+\log({\rm O/H})\sim8.5$ 
\citep[or $Z\sim0.7\ Z_\odot$;][]{martinez25} and
M4327: $12+\log({\rm O/H})\sim8.15$ 
\citep[or $Z\sim0.3\ Z_\odot$;][]{curti25b}; and 
(3) the WR \ion{N}{3} \W4642 line is much stronger in RXCJ2248-ID3, which  
lacks WR \ion{C}{4} \W\W5803,5814 emission, while the Sunburst Arc and M4327 
exhibit both \ion{N}{3} and \ion{C}{4} emission.
These differences support a scenario in which the $z\sim2$ WR galaxies hosts both 
WC and WN stars, but the more metal-poor RXCJ2248-ID3 hosts a young population 
of WN stars with no or very little WC contribution.


\subsection{Relative Chemical Enrichment from WN Stars}\label{sec:chemevol}
With the highest redshift detection of WN stars to date, we now explore 
their chemical yields as a source of the abundance pattern in RXCJ2248-ID3.
\citet{charbonnel23} performed a comparable analysis for the extreme N/O 
ratio observed in GN-z11 and found that such rapid nitrogen enrichment 
could arise from normal massive stars with $M_\star \sim 20-120\,M_\odot$ or 
from supermassive stars ($M_\star\gtrsim1000\,M_\odot$) in protoglobular 
cluster environments.
Their results and those of \cite{marques-chaves24} further demonstrated that 
the short-lived WN-like phase can produce large N/O ratios within a few Myr
of the burst, 
consistent with the timescales inferred here, but that the observed C/O ratios 
are only compatible over a very short time interval.
Building on this theoretical groundwork, \citet{kobayashi24} showed that
a dual-burst chemical evolution model with a short WR-dominated enrichment phase 
could also match GN-z11's enrichment pattern.
Similarly, \citet{marques-chaves24} used N yields from rotating massive stars to
demonstrate that a young, WR-dominated stellar population could reproduce 
the observed CNO enrichment pattern in CEERS-1019.

The models above provide a valuable physical framework for linking stellar yields to 
galaxy-scale abundance evolution at early times.
To extend the methodologies outlined by these works to RXCJ2248-ID3, 
we first examine the dual-burst chemical evolution model of 
\citet{kobayashi24}, which was fine-tuned to reproduce the enhanced N/O in GN-z11 
\citep[reported by][]{maiolino23}. 
This model invokes two bursts of star formation, where the second triggers a 
narrow ($\lesssim1$ Myr) phase of WR-dominated enrichment. 
While the model can easily reach the N/O enrichment level of RXCJ2248-ID3,
it was also designed to yield the higher O/H and C/O abundances observed in
GN-z11 than in RXCJ2248-ID3, which was achieved, in part, by enrichment from
WC stars.
The updated O/H abundance for GN-z11 determined by \citet{martinez25} makes it 
consistent with some carbon enrichment from WC stars, as shown in Figure~\ref{fig:WCN}. 
However, with a metallicity of only $Z\sim0.10\ Z_\odot$, the WR population in 
RXCJ2248-ID3 is expected to consist of few WC stars, and so an updated chemical 
evolution model is needed to match its unique CNO abundance pattern.

We modify the \citet{kobayashi24} dual-burst model to be more appropriate for 
the metal-poor conditions in RXCJ2248-ID3.
In particular, the galactic chemical evolution (GCE) model uses the same star 
formation history and the standard IMF (for 0.01--120$M_\odot$) as in the 
fiducial model in \citet{kobayashi24} but reduces the contribution from WC stars. 
C/O ratios of the nucleosynthesis yields vary depending on the uncertain nuclear 
reaction rates (e.g., $^{12}$C($\alpha$,$\gamma$)$^{16}$O) and the treatment of 
convection and mass loss \citep[][]{kobayashi06}. 
In the updated model, $^{12}$C and $^{16}$O yields are taken from 
\citet{kobayashi20} for all mass ranges of stars but the contributions from the WC 
wind phase is scaled to $\sim15\%$ in order to match the empirical trends and 
theoretical expectations that most massive stars will have have insufficient 
winds to remove their He envelopes at such low metallicities. 

We plot the updated metal-poor dual-burst model in the bottom row of 
Figure~\ref{fig:CNO} as a time-series of points that are color coded by the 
age since the onset of the second burst.
In this model, the observed N/O, O/H, and C/O abundances of RXCJ2248-ID3 are 
reached simultaneously $\sim4.2$ Myr after the onset of the second burst.
This young age is consistent with enrichment from WN stars and with the 
derived clump age of $1.6^{+11.9}_{-0.9}$ Myr \citep[][]{claeyssens26}.
Thus, the N/O-enhanced and relatively C/O-deficient conditions in RXCJ2248-ID3 
are produced by a short-lived evolutionary phase following intense, 
bursty star formation.

We note that the duration and impact of the WN phase may be significantly 
extended if the stars evolve in binary systems. 
In the \citet{limongi18} single-star models, the WN phase typically lasts 
$\sim0.03-0.3$ Myr and, due to the metallicity-dependent winds, require high 
initial masses ($\sim40\ M_\odot$) to expose the He- and N-rich layers. 
However, in close binaries, envelope stripping via mass transfer or common-envelope 
evolution can induce WR phases in lower-mass stars ($20-30\ M_\odot$), 
largely independent of the stellar metallicity.
This channel can significantly prolong the WN lifetime (up to $\sim1$ Myr) depending 
on the binary mass ratio and separation 
\citep[e.g.,][]{eldridge17,gotberg19,aguilera-dena22}. 
As a result, binary evolution may enhance both the frequency and duration of 
the chemically-selective N/O enrichment phase, such as that observed in RXCJ2248-ID. 
On the other hand, \citet{boco25} successfully modeled observations of single WR 
stars in the SMC, suggesting that binary stripping may not be required to produce WR 
stars at low metallicity.
Clearly, the frequency, lifetimes, and formation channels of WR stars in 
low-metallicity environments are not yet well understood. 
Future work incorporating current binary and single star WR pathways into chemical 
evolution models may, therefore, be essential for capturing the full range of 
nitrogen feedback in low-metallicity starbursts at high-redshift.

Taken together, the massive star enrichment scenario presented here, and 
explored in \citet{charbonnel23}, \citet{marques-chaves24}, and 
\citet{kobayashi24}, demonstrate that selective enrichment of nitrogen by 
WN-dominated feedback can naturally reproduce the observed CNO abundance 
pattern in compact, low-metallicity starbursts such as RXCJ2248-ID3.
We can now paint a full picture of the ISM in RXCJ2248-ID3.
The consistency of N/O across ions spanning a wide range of ionization potentials 
suggests that the WN-enriched material has been efficiently mixed throughout 
the ionized gas. 
This apparent chemical homogeneity does not contradict the strong density and temperature 
stratification inferred from our diagnostics: a clumpy or multiphase ISM can remain 
compositionally uniform if the enriched ejecta are well dispersed. 
Given the extreme compactness of RXCJ2248-ID3 ($R_e\approx20$ pc), the characteristic 
dynamical and sound-crossing times are only a few $\times 10^5$ yr, comparable to 
or shorter than the duration of the WN phase itself. 
Under such conditions, turbulent and radiative mixing can rapidly homogenize the 
heavy-element yields, producing a chemically uniform yet physically structured nebula.

\subsection{The N Mass Budget}\label{sec:Nmass}
A crucial point of validation is whether an intense burst of star formation 
so early in the Universe 
could have produced the {\it amount} of N present in RXCJ2248-ID3.
Similar to the analysis in \citet{marques-chaves24}, we test this by first estimating 
the ionized nitrogen mass using:
\begin{equation}
    M_{\rm N_{ion.}} = M_{\rm H_{ion.}}\left(\frac{{\rm N}}{\rm H}\right)_{\rm ion.}
    \left(\frac{m_{\rm N}}{m_{\rm H}}\right),
\end{equation}
where the atomic mass ratio is $m_{\rm N}/m_{\rm H}=14$ and N/H is the nitrogen 
abundance of the ionized gas.
The hydrogen gas mass $M_{\rm H}$ is derived from the H$\alpha$ luminosity:
\begin{equation}
    M_{\rm H} = \frac{m_{\rm H}L_{\rm H\alpha}}{h\nu_{\rm H\alpha}\alpha_{\rm H\alpha}^{\rm eff}n_e} 
    \sqrt{\epsilon},
\end{equation}
where $m_{\rm H}=1.67\times10^{-27}$ kg, 
$h=6.626\times10^{-27}$ erg$\cdot$s, 
$\nu_{\rm H\alpha}$ is the frequency of the H$\alpha$ emission line, and 
$\alpha_{\rm H\alpha}^{\rm eff}=6.078\times10^{-14}$ cm$^{3}$ s$^{-1}$ is the 
Case B effective recombination coefficient for H$\alpha$ assuming a 
$T_e=\thigh\times10^4$ K.
We estimate the H$\alpha$ luminosity using a luminosity distance of 
$d_L(z=6.1025)=1.817\times10^{29}$ cm, the collision-corrected narrow component 
H$\alpha$ flux, and a magnification of $\mu=6.8877$ \citep{furtak26} to be
$L_{\rm H\alpha}=~$\LHA\ erg s$^{-1}$. 
Combining this $L_{\rm H\alpha}$ with the EW(H$\alpha$)~=~\EWHA\,\AA, we 
derive the star formation rate (SFR) using the simulation-based SFR(H$\alpha$) 
calibration from \citet{kramarenko25}.
This method was developed to be more appropriate for the bursty conditions at 
high-redshift than traditional calibrations and gives a 
SFR~$=\SFR\ M_\odot$ yr$^{-1}$, is similar to the SED-derived 
SFRs assuming a constant SFH for 1 Myr ($4.7\ M_\odot$ yr$^{-1}$) and 
10 Myr ($4.1\ M_\odot$ yr$^{-1}$; see Table~\ref{tbl1}).
Adopting a filling factor of $\varepsilon=0.01-$$0.10$, assuming a compact 
starburst \citep[e.g.,][]{kennicutt84,stasinska97}, a density of $10^4$ 
cm$^{-3}$, and the measured N/H value, we calculate the ionized nitrogen mass 
to be $M_{\rm N_{ion.}} \approx 18.2-57.5\ M_\odot$.

We then compute the total nitrogen mass that can be produced 
by the recent burst of star formation 
using  the integrated nitrogen yield produced by the modified \citet{kobayashi24} 
model for the SED-derived stellar 
mass of $M_\star = 1.96\times10^7\ M_\odot$ assuming a continuous SFH
over the duration of the second burst ($\sim4.2$ Myr).
This results in a total N mass of $435\ M_\odot$, implying that $\sim4-13\%$ 
of the gas is retained from the WN winds and ionized when matched to the expected 
ionized N mass ($ 18.2-57.5\ M_\odot$) from the crudely calculated observed 
value. 
Thus, WN stars formed in a recent burst within a compact, high-density, 
and very clumpy / inhomogeneous (low-filling-factor) environment can plausibly 
explain the N mass in RXCJ2248-ID3, even at low metallicity ($\sim10\%\ 
Z_\odot$), without invoking a top-heavy IMF or exotic enrichment channels.

\subsection{The Ephemeral Imprint of WN Star on High\texorpdfstring{$-z$}{z} Galaxies}\label{sec:time}
\begin{figure}
\begin{center}
	\includegraphics[width=1.0\linewidth,trim=0mm 0mm 10mm 10mm,clip]{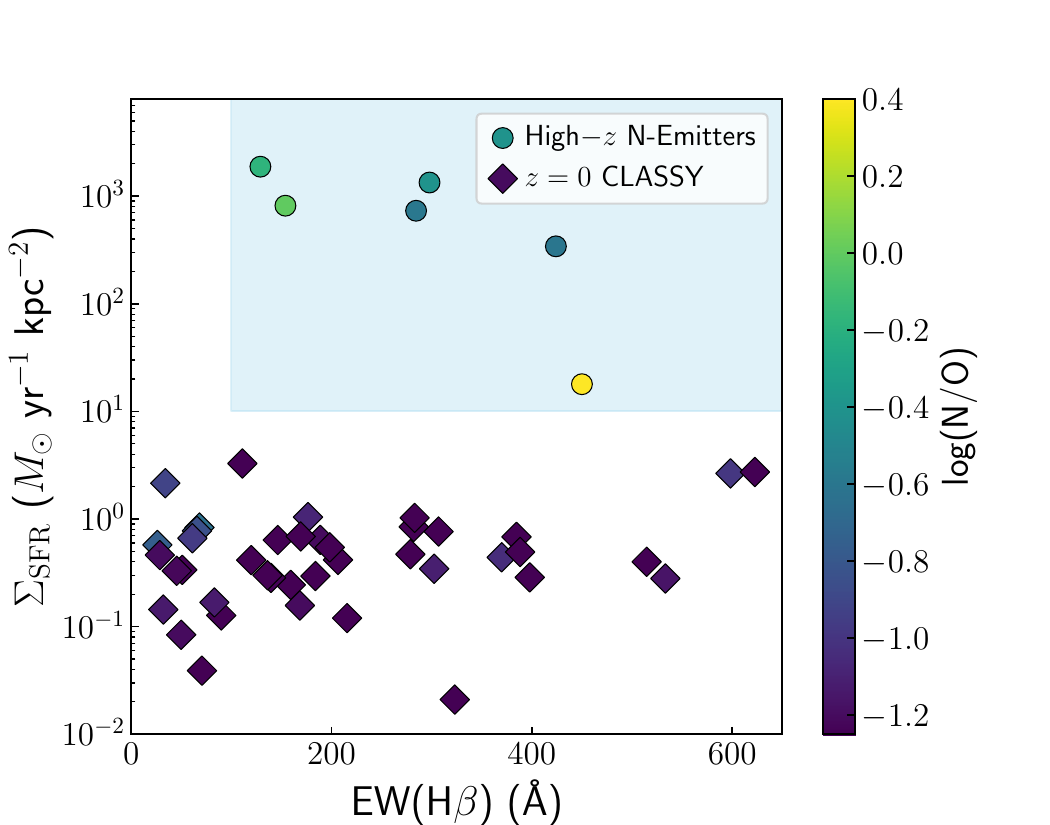}
\caption{
Star formation rate surface density versus H$\beta$ equivalent width for
high-redshift ($z>5$) N-emitters versus $z\sim0$ galaxies 
from the CLASSY survey (SFR: \citealt{berg22}; N/O: \citealt{arellano-cordova25}), 
which have enhanced SFRs similar to $z\sim2-3$ galaxies.
High-redshift N-emitters are only observed at young ages ($\lesssim5$ Myr), 
as indicated by the high H$\beta$ equivalent widths (EW$>200$ \AA), and 
in compact, dense environments 
($\Sigma_{\rm SFR}>10\ M_\odot{\rm\ yr^{-1}\ kpc^{-1}}$).
Note that RXCJ2248-ID3 is plotted here using the properties derived 
from \citet{topping24a} for continuous star formation to be consistent
with the other N-emitter measurements.
}\label{fig:SFRvEW}
\end{center}
\end{figure}

The prominence of N/O enhancement at $z\gtrsim5$ but 
relative rarity in local star-forming galaxies likely reflects a combination of 
environmental conditions and evolutionary factors that are unique to the early 
Universe. 
To examine the likely environments, we plot the SFR surface density 
($\Sigma_{\rm SFR}$) versus EW of H$\beta$ in Figure~\ref{fig:SFRvEW} for both 
$z\gtrsim6$ N emitters (RXCJ2248-ID3: \citealt{topping24a}, this work;
GNz9p4: \citealt{schaerer24};
GNz11, EW(H$\beta$) inferred from H$\gamma$: \citealt{bunker23,tacchella23};
GDS 3073: \citealt{vanzella10,ubler23,ji24};
CEERS-1019: \citealt{larson23,marques-chaves24};
A1703-zd6: \citealt{topping25a})
and local star-forming galaxies with enhanced SFRs from 
the COS Legacy Archive Spectroscopic SurveY 
(CLASSY; \citealt{berg22,james22}; N/O from \citealt{arellano-cordova25}).
The high-redshift galaxies, such as RXCJ2248-ID3, exhibit compact morphologies 
($R_e\lesssim10^2$ pc) that lead to much higher SFR surface densities than 
seen at $z\sim0$, as well as bursty star formation histories that favor the rapid 
buildup of massive stars capable of entering the short-lived WN phases 
($M_\star>20\ M_\odot$). 
The high-redshift N emitters also have high H$\beta$ EWs ($>200$ \AA) that are 
indicative of young current bursts of star formation ($<5$ Myr).
This suggests that compactness alone is not enough to observe enhanced N/O; 
we must also observe these galaxies at the fleeting moments of very young bursts
when WR stars are most active.


Figure~\ref{fig:SFRvEW} suggests a scenario of elevated N/O at low metallicity being 
preferentially seen in galaxies with high SFR surface densities and young stellar 
ages \citep[e.g.,][]{topping24a,schaerer24,martinez25}.
\citet{marques-chaves24} also suggest that the elevated N/O and high-ionization 
spectrum of CEERS-1019 trace a short evolutionary window of a $\lesssim5$\,Myr burst 
dominated by WN-like feedback.
Furthermore, the theoretical models of \citet{charbonnel23} predict that such phases 
are characteristic of young, dense stellar systems, potentially analogous to 
protoglobular clusters, reinforcing that our observed WN-driven enrichment is a 
natural outcome of clustered, bursty star formation at early times.

At low metallicity, weaker stellar winds require higher initial masses for stars to 
reach the WR phase, so a larger total stellar mass must form in a burst to produce a 
detectable population of WN stars. 
In compact galaxies beyond cosmic noon, this condition is naturally met in systems with 
high star formation rate surface densities, which statistically sample the upper 
IMF more fully and produce a detectable population of WN stars 
\citep[e.g.,][]{brinchmann08,shirazi12}.
Furthermore, the WR enrichment signature is short-lived: 
it must be captured during the brief WN-dominated phase ($t_{\rm burst}\lesssim5$ Myr and 
$\Delta t_{\rm WN}\lesssim0.3$ Myr), before dilution from WC stars, CCSNe, 
or delayed AGB enrichment. 
These timing constraints imply that only a small fraction of the star-forming 
galaxies in the distant Universe will be caught in this phase. 
The detection of WR-driven N/O enhancement at high-redshift thus reflects a brief 
evolutionary stage where intense, rapid feedback from a large number of WN stars 
briefly imprints nonuniform elemental enrichment patterns (i.e., elevated N/O) 
that are expected to be quickly washed away. 
As soon as the system evolves beyond the WN phase, subsequent WC or CCSN yields 
will rapidly dilute the N excess and alter the overall abundance patter 
(e.g., increasing C/O, lowering N/C).

Recently, \citet{topping25a} showed that galaxies with significant \ion{N}{4}] 
emission (corresponding to extreme N/O enhancement), are found exclusively  
amongst galaxies with extreme [\ion{O}{3}]+H$\beta$ equivalent widths (EW) of 
$2600-4200$ \AA.
Galaxies with such high [\ion{O}{3}]+H$\beta$ EWs are in the upper 2\% tail of the EW 
distribution at $z\gtrsim4$ and are outliers at $z\sim0$.
This strongly suggests that high N/O outliers are confined to the youngest stellar 
populations undergoing their most intense bursts of star formation in the early Universe 
\citep[e.g.,][]{endsley23,matthee23,endsley25,topping25a}. 

In this context, \citet{topping25a} found that 30\% of galaxies with 
EW$_{\rm [OIII]+H\beta}>2000$ \AA\ show strong nitrogen emission, corresponding to 
$\sim$0.6\% of their UV-selected parent population. 
If this 0.6\%\ population corresponds to enhanced N/O during the 
$\Delta t_{\rm WN}\sim0.3$ Myr WN phase, it would imply a characteristic burst 
timescale of $\sim50$ Myr. 
A practical consequence is that young bursts substantially increase the 
light-to-mass ratios and, thus, the likelihood of detection in flux-limited samples
\citep[e.g.,][]{sun23,mason23,munoz23}. 
Therefore, the observed frequency of strong nitrogen emitters at fixed $M_{\rm UV}$ 
is likely biased high relative to their intrinsic abundance (e.g., at fixed stellar mass).
Given the detectability bias toward burst phases, this $t_{\rm burst}$ may represent 
a lower limit, with the true interval plausibly longer. 
This timescale is supported by recent analyses of the scatter in the star-forming
main sequence and time-resolved SFR indicators at $z\sim3-9$ that suggest burst cycles 
of tens-of‐Myr timescales \citep[albeit with broad distributions, e.g.,][]{simmonds25}.
Thus, the combination of extreme-EW selection and \ion{N}{4}] frequency provides a 
novel timing argument that WN-driven enrichment is tightly coupled to very young, 
transient starburst phases beyond cosmic noon.

Taken together, the arguments presented in this work suggest that nitrogen outliers 
are not exotic exceptions, but rather a brief, WN-enriched phase that any high-redshift 
galaxy with sufficiently high SFR surface density can pass through. 
In contrast, numerous low-redshift WR galaxies exhibit young populations that include 
WN and WC stars but show little or no N/O enhancement 
\citep[e.g.,][]{izotov06b,kehrig13}. 
This difference underscores that similar stellar populations do not guarantee the 
same chemical signatures; instead, the extreme densities, compactness, and rapid 
mixing timescales of high-redshift starbursts likely make WN-driven enrichment 
both more pronounced and more transient.
In this view, N/O outliers in the early Universe are not anomalies, but rather are the 
chemical fingerprints of galaxies caught midburst, showing fleeting yet inevitable 
markers of early galaxy evolution.


\section{Conclusions}\label{sec:conclusions}
We have presented a detailed enrichment scenario by WN stars that explains the extreme 
nitrogen enrichment in the metal-poor ($\sim10\%\ Z_\odot$), 
high surface-density ($1.34\times10^3\ M_\odot$ pc$^{-2}$), high-redshift 
($z=6.1025$), lensed galaxy RXCJ2248-ID3.
These measurements were made possible by exceptionally deep JWST/NIRSpec 
medium-resolution spectroscopy of RXCJ2248-ID3, 
obtained as part of the GLIMPSE-D survey. 
The unprecedented depth and S/N of the GLIMPSE-D spectrum allow spectral measurements 
typically limited to the nearby Universe, including consistent broad components in 
the Balmer series and [\ion{O}{3}] \W4364 and \W\W4960,5008 lines, 
faint [\ion{Ar}{4}] \W\W4713,4741 emission, and signatures of WR stars. 
Specifically, we detected emission characteristic of WN-type stars, 
including strong \ion{N}{3} \W4642 and broadened \ion{He}{2} \W1640 and \W4687 emission, 
marking RXCJ2248-ID3 as the most distant galaxy to date with spectroscopic detections of 
WR stars. 

We performed a detailed nebular analysis, self-consistently measuring the 
reddening, high-ionization temperature ($T_e({\rm O}^{+2})$), 
and densities from five different diagnostics across a wide ionization range.
We measure a low reddening value of $E(B-V)=\ebv^{+\ebveu}_{-\ebved}$ from the 
H$\gamma$/H$\beta$ ratio, but find an excess in the H$\alpha$/H$\beta$ ratio
of \HAexc\ due to collisional excitation of H$\alpha$.
The measured densities span the range of 
$\nSII\times10^3\ {\rm cm}^{-3}\leq n_e \leq \nNIV\times10^5\ {\rm cm}^{-3}$ and 
show strong evidence for nebular density stratification, 
with systematically higher densities in the highest-ionization gas and 
UV emission tracing gas at higher densities than those traced by optical diagnostics. 
This structure implies a highly clumpy, multiphase ISM.
We note that such high-density, multiphase gas leads to densities from 
optical diagnostics that are biased to the low end of the density range 
due to their low critical densities.
Therefore, we recommend using UV density diagnostics because they are more 
robust in high-density environments: 
$n_{e}$(Si$^{+2}$), $n_{e}$(C$^{+2}$), and $n_{e}$(N$^{+3}$) trace the 
densities in the low-, intermediate-, and high-ionization gas, respectively.
As a result, we measure a direct-method metallicity of 
$12+\log(\rm O/H)=\loh\pm\lohe$. 

Using the full rest-UV+optical spectra, we present the first robust, 
consistent measurements of N/O abundance in any galaxy using three 
ionization stages of nitrogen (N$^+$/O$^+$, N$^{+2}$/O$^{+2}$, N$^{+3}$/O$^{+2}$).
The uniformity of our N/O measurements suggests that the N/O enrichment 
is spatially extended and well mixed throughout the ionized ISM.
Empirical trends suggest C/O should follow a similar trend as N/O and,
thus, also be enhanced.
In contrast, we find  C/O to be significantly depleted relative to N/O, 
suggesting nonuniform elemental enrichment likely driven by WN stars with 
little to no contribution from WC stars.

The CNO abundance pattern is best reproduced by a modified version of the 
dual-burst chemical evolution model from \citet{kobayashi24} that reduces the
contribution from WC stars relative to WN stars, as expected in metal-poor 
environments. 
The resulting short-lived WN phase ejects N-rich, C-poor material. 
We use this chemical evolution model to assess whether the observed 
N mass can plausibly arise from the recent star formation in RXCJ2248-ID3
and estimate an ionized N mass of $435\ M_\odot$. 
This value is consistent with the N mass estimated from the observed 
emission lines of $18.2-57.5 M_\odot$ if $4-13\%$  of the N gas is
ionized. 

These results demonstrate that standard stellar evolution models can reproduce 
both the CNO pattern and the total nitrogen mass observed without invoking an 
exotic IMF or enrichment channel. 
The uniform N/O ratios across multiple ionization zones further suggest that the 
WN yields were rapidly mixed into a relatively pristine ambient ISM, preserving 
the global enhancement observed in RXCJ2248-ID3.
Although RXCJ2248-ID3 exhibits strong density and temperature stratification, 
this structural complexity does not necessarily imply chemical inhomogeneity. 
The consistent N/O ratios across ions tracing vastly different physical conditions 
indicate that the enriched material was efficiently dispersed throughout the 
multiphase ISM. 
In such a compact ($R_e\approx20$ pc), high-pressure environment, turbulent and 
radiative mixing can homogenize the chemical composition on timescales comparable to, 
or shorter than, the brief WN phase itself, yielding a chemically uniform yet 
physically clumpy nebula.\looseness=-2

Importantly, the abundance pattern and physical conditions observed in RXCJ2248-ID3 
can only be explained if the galaxy is caught during a narrow evolutionary window
within a few Myr of a massive, compact starburst when WN stars dominate chemical 
feedback. 
At low metallicity, stars require higher initial masses to reach the WR phase, 
making such enrichment episodes rare and dependent on sufficiently high SFRs to 
fully populate the upper IMF. 
Furthermore, the WN phase itself is extremely short-lived ($\sim0.03-0.3$ Myr), 
and easily masked by subsequent WC winds, CCSNe, or AGB stars contributions. 
These timing and SFR constraints make WR-driven N/O enhancement a rare phenomenon 
associated with extreme starburst conditions that are more common in the early Universe, 
and which are scarce in the local Universe. 

Our results suggest that the WN-driven N/O enrichment we observe is not a peculiar 
property of a single system, but rather a brief phase that essentially all 
high-redshift galaxies ($z>5$) with sufficiently high SFR surface densities to produce 
significant numbers of WN stars likely undergo. 
In particular, the work of \citet{topping25a} can be used to link N/O outliers 
to the most extreme [\ion{O}{3}]+H$\beta$ EWs.
The observed frequency of such EWs combined with the short lifetime 
of the WN phase implies a burst cycle of order $\sim50$ Myr, consistent with 
galaxies repeatedly cycling through short, bursty episodes of enrichment. 
Thus, the GLIMPSE-D spectrum of RXCJ2248-ID3 provides not only the first direct 
evidence of WN stars shaping the chemical evolution of $z>5$ galaxies, but also 
a timing argument that situates N/O outliers as a natural, fleeting, phase of 
high-redshift star formation.

Taken together, our findings are a glimpse into a short-lived phase of 
chemically-selective enrichment from WN stars at cosmic dawn, providing a 
physically self-consistent solution to the extreme N/O enhancement and 
relative C/O depletion observed in RXCJ2248-ID3 and galaxies like it.
Thus, RXCJ2248-ID3 serves as a benchmark case for interpreting chemically 
enriched, stratified, multiphase starbursts in the early Universe. 

\facilities{JWST (NIRSpec)}
\software{
\texttt{astropy} \citep{astropy:2013, astropy:2018, astropy:2022},
\texttt{calwebb},
\texttt{jupyter} \citep{kluyver16},
\texttt{numpy} version 1.26 \citep{harris20},
\texttt{PyNeb} version 1.1.14 \citep{luridiana15},
\texttt{python},
\texttt{lmfit} \citep{newville15}, 
\texttt{mpfit} \citep{markwardt09},
\texttt{scipy.optimize.curve\_fit},
\texttt{BAGPIPES} \citep{carnall18},
\texttt{BPASS} version 2.14 \citep{eldridge17},
\texttt{cloudy} version 23.01 \citep{chatzikos23, gunasekera23},
\texttt{msaexp} pipeline version 0.9.8 \citep{msaexp}
}

\begin{acknowledgements}
We thank the referee for their thorough review of our calculations and analysis
and for their helpful suggestions, which greatly improved the robustness of our results
and clarity of the text. 
This work is based on observations made with the NASA/ESA/CSA James Webb Space Telescope. 
The data were obtained from the Mikulski Archive for Space Telescopes at the Space Telescope Science Institute, which is operated by the Association of Universities for Research in Astronomy, Inc., under NASA contract NAS 5-03127 for JWST. 
These observations are associated with program \#9223.
This work has received funding from the Swiss State Secretariat for Education, Research and Innovation (SERI) under contract number MB22.00072, as well as from the Swiss National Science Foundation (SNSF) through project grant 200020\_207349. The Cosmic Dawn Center (DAWN) is funded by the Danish National Research Foundation under grant DNRF140.
The Dunlap Institute is funded through an endowment established by the David Dunlap family and the University of Toronto.
We acknowledge the support of the Canadian Space Agency (CSA) [25JWGO4A06]. HA acknowledges support from CNES, focused on the JWST mission, and the Programme National Cosmology and Galaxies (PNCG) of CNRS/INSU with INP and IN2P3, co-funded by CEA and CNES and support by the French National Research Agency (ANR) under grant ANR-21-CE31-0838.

The JWST data presented in this article from program \#9223 were obtained from 
the Mikulski Archive for Space Telescopes (MAST) at the Space Telescope Science 
Institute. 
The specific observations analyzed can be accessed via 
\dataset[10.17909/8642-1k68]{https://doi.org/10.17909/8642-1k68}.
\end{acknowledgements}
\clearpage

\bibliographystyle{aasjournal}
\bibliography{mybib}

\clearpage

\end{document}